\documentclass[twocolumn]{aastex63}
\pdfoutput=1 
\usepackage[T1]{fontenc}
\usepackage{amsmath,amstext}
\usepackage{amssymb}
\usepackage[T1]{tipa} % To add new commands
\usepackage{apjfonts} 
\usepackage[figure,figure*]{hypcap}
\usepackage{microtype}
\usepackage{tablefootnote}
\usepackage{booktabs}  
\usepackage{longtable}
\usepackage{hyperref}
\usepackage{CJKutf8}

\newcommand{\kkoname}{k'ni\textipa{P}atn k'l$\left._\mathrm{\smile}\right.$stk'masqt}

\begin{document}
\begin{CJK*}{UTF8}{gbsn}
\shortauthors{Dong et al.}

\shorttitle{Optical Transients and FRBs}

\title{Searching for Historical Extragalactic Optical Transients Associated with Fast Radio Bursts}

\newcommand{\NU}{\affiliation{Department of Physics and Astronomy, Northwestern University, Evanston, IL 60208, USA}}

\newcommand{\CIERA}{\affiliation{Center for Interdisciplinary Exploration and Research in Astronomy (CIERA), Northwestern University, 1800 Sherman Avenue, Evanston, IL 60201, USA }}

\newcommand{\MU}{\affiliation{Department of Physics, McGill University, 3600 rue University, Montr\'eal, QC H3A 2T8, Canada}}

\newcommand{\TSI}{\affiliation{Trottier Space Institute, McGill University, 3550 rue University, Montr\'eal, QC H3A 2A7, Canada}}

\newcommand{\UVA}{\affiliation{Anton Pannekoek Institute for Astronomy, University of Amsterdam, Science Park 904, 1098 XH Amsterdam, The Netherlands}}

\newcommand{\DI}{\affiliation{Dunlap Institute for Astronomy and Astrophysics, 50 St. George Street, University of Toronto, ON M5S 3H4, Canada}}

\newcommand{\DAA}{\affiliation{David A. Dunlap Department of Astronomy and Astrophysics, 50 St. George Street, University of Toronto, ON M5S 3H4, Canada}}

\newcommand{\UCSC}{\affiliation{Department of Astronomy and Astrophysics, University of California, Santa Cruz, 1156 High Street, Santa Cruz, CA 95060, USA}}

\newcommand{\SKAO}{\affiliation{SKA Observatory, 26 Dick Perry Ave, Kensington WA 6151 Australia}}

\newcommand{\WVUPHAS}
{\affiliation{Department of Physics and Astronomy, West Virginia University, PO Box 6315, Morgantown, WV 26506, USA }}

\newcommand{\WVUGWAC}
{\affiliation{Center for Gravitational Waves and Cosmology, West Virginia University, Chestnut Ridge Research Building, Morgantown, WV 26505, USA}}

\newcommand{\MITK}
{\affiliation{MIT Kavli Institute for Astrophysics and Space Research, Massachusetts Institute of Technology, 77 Massachusetts Ave, Cambridge, MA 02139, USA}}

\newcommand{\MITP}
{\affiliation{Department of Physics, Massachusetts Institute of Technology, 77 Massachusetts Ave, Cambridge, MA 02139, USA}}

\newcommand{\LAM}
{\affiliation{Laboratoire d'Astrophysique de Marseille, Aix-Marseille Univ., CNRS, CNES, Marseille, France}}

\newcommand{\PI}
{\affiliation{Perimeter Institute of Theoretical Physics, 31 Caroline Street North, Waterloo, ON N2L 2Y5, Canada}}

\newcommand{\YORK}
{\affiliation{Department of Physics and Astronomy, York University, 4700 Keele Street, Toronto, ON MJ3 1P3, Canada}}

\newcommand{\MIBR}
{\affiliation{Miller Institute for Basic Research, University of California, Berkeley, CA 94720, United States}}

\newcommand{\UCBASTRO}
{\affiliation{Department of Astronomy, University of California, Berkeley, CA 94720, United States}}

\author[0000-0002-9363-8606]{Y.~Dong (董雨欣)}
\CIERA
\NU

\author[0000-0002-5740-7747]{C.~D.~Kilpatrick}
\CIERA

\author[0000-0002-7374-935X]{W.~Fong}
\CIERA
\NU

\author[0000-0002-8376-1563]{A.P.~Curtin}
\MU
\TSI

\author{S.~Opoku}
\NU

%%  Alphabetical list   
\author[0000-0001-5908-3152]{B.~C.~Andersen}
\MU
\TSI

\author[0000-0001-6422-8125]{A.~M.~Cook}
\MU
\TSI
\UVA

\author[0000-0003-0307-9984]{T.~Eftekhari}
\CIERA
\NU

\author[0000-0001-8384-5049]{E.~Fonseca}
\WVUPHAS
\WVUGWAC

\author[0000-0002-3382-9558]{B.~M.~Gaensler}
\UCSC
\DAA
\DI

\author[0000-0003-3457-4670]{R.~C.~Joseph}
\MU
\TSI

\author[0000-0003-4810-7803]{J.~F. Kaczmarek}
\SKAO

\author[0009-0007-5296-4046]{L.~A.~Kahinga}
\UCSC

\author[0000-0001-9345-0307]{V.~Kaspi}
\MU 
\TSI

\author[0000-0003-2116-3573]{A.~E.~Lanman}
\MITK
\MITP

\author[0000-0002-5857-4264]{M.~Lazda}
\DAA
\DI

\author[0000-0002-4209-7408]{C.~Leung}
\MIBR
\UCBASTRO

\author[0000-0002-4279-6946]{K.~W.~Masui}
\MITK
\MITP

\author[0000-0002-2551-7554]{D.~Michilli}
\LAM

\author[0000-0003-0510-0740]{K.~Nimmo}
\MITK

\author[0000-0002-8897-1973]{A.~Pandhi}
\DAA
\DI

\author[0000-0002-8912-0732]{A.~B.~Pearlman}
\altaffiliation{Banting Fellow, McGill Space Institute~(MSI) Fellow, \\ and FRQNT Postdoctoral Fellow.}
\MU
\TSI

\author[0000-0002-4623-5329]{M.~Sammons}
\MU
\TSI

\author[0000-0002-7374-7119]{P.~Scholz}
\YORK
\DI

\author[0000-0002-4823-1946]{V.~Shah}
\MU
\TSI

\author[0000-0002-6823-2073]{K.~Shin}
\MITK
\MITP

\author[0000-0002-2088-3125]{K.~Smith}
\PI

\begin{abstract}
% keep under or at 250 words
We present a systematic search for past supernovae (SNe) and other historical optical transients at the positions of fast radio burst (FRB) sources to test FRB progenitor systems. Our sample comprises 83 FRBs detected by the Canadian Hydrogen Intensity Mapping Experiment (CHIME) and its \kkoname~(KKO) Outrigger, along with 93 literature FRBs representing all known well-localized FRBs. We search for optical transients coincident in position and redshift with FRBs and find no significant associations within the 5$\sigma$ FRB localization uncertainties except for a previously identified potential optical counterpart to FRB\,20180916B. By constraining the timescale for SN ejecta to become transparent to FRB emission, we predict that it takes at least 6--10 years before the FRB emission can escape. From this, we infer that $\approx\!7\%$ of matched optical transients, up to $30\%$ of currently known SNe, and up to $40\%$ of core-collapse SNe could have an observable FRB based on timescales alone. We derive the number of new, well-localized FRBs required to produce one FRB-SN match by chance, and find it will take $\sim 22,700$ FRBs to yield one chance association at the projected CHIME/FRB Outrigger detection rate. Looking forward, we demonstrate redshift overlap between SNe detected by the upcoming Vera C. Rubin Observatory and CHIME/FRB Outrigger FRBs, indicating the prospect of an increase in potential associations at redshift $z < 1$. Our framework is publicly available, flexible to a wide range of transient timescales and FRB localization sizes, and can be applied to any optical transient populations in future searches.

\end{abstract}


\section{Introduction}
\label{sec:intro}

% Explain FRBs, what they are and the progenitor problem: Not fully known, but the prevailing model is a magneter engine because of the coherent radio emission, energy required, and the connection to a Galactic magnetar. 
Fast radio bursts (FRBs) are luminous, millisecond-duration radio pulses that primarily originate at cosmological distances \citep{Lorimer07, Thornton2013, Bannister17, Cordes19}. Despite the discovery of over 800 FRB sources published to date (e.g., \citealt{CHIMEcat1, law2024, Shannon24}), the physical origin(s) of FRBs remain elusive. While a definitive consensus has yet to be reached on FRB progenitors, the leading theory invokes magnetars whose strong magnetic field strengths can produce the bursts through magnetospheric processes or relativistic shocks \citep{Metzger18,Lyutikov19,Wadiasingh19,Lu20,Zhang22,Mckinven25,Nimmo25}. This model is largely motivated by the extreme energy densities required for short coherent radio emission down to nanosecond timescales (e.g., \citealt{Nimmo22}). Bolstering this scenario, FRB\,20200428D was unambiguously associated with the Galactic magnetar SGR\,J1935$+$2154 \citep{Bochenek20, CHIME2020}. 

The association between FRB\,20200428D and SGR\,J1935$+$2154 provides clear evidence that magnetars are responsible for at least some FRBs. However, such a direct association between FRBs and magnetars is currently only feasible within the Milky Way (MW) where most known magnetars reside; in contrast, all FRBs except for FRB\,20200428D are extragalactic. Furthermore, while there exist arguments for multiple formation pathways of magnetars capable of producing FRBs \citep{Totani13, Metzger17, Margalit19, WangFY20, Zhong20, Kremer21, Kirsten22,  Rao25}, it is not clear how strongly each formation pathway contributes to the total FRB rate. An alternative and powerful approach is to associate FRBs with other types of multi-wavelength and/or multi-messenger counterparts that have well-established progenitor systems. Indeed, this kind of direct test has been applied to searches for gravitational wave events and short gamma-ray bursts as potential counterparts to FRBs \citep{Wang22GW, Moroianu23, Curtin24}. In these cases, the FRB progenitor is hypothesized to form through binary neutron star (BNS) mergers with long delay times. In parallel, search efforts have been dedicated to identifying long gamma-ray bursts and superluminous supernovae (SLSNe) as signatures of FRB progenitors \citep{Eftekhari19, Eftekhari21, law19, Curtin23, Curtin24}. These events are thought to arise through a prompt channel relative to star formation, in which a young, massive star collapses directly into a magnetar, producing the transient shortly afterward.

Another illuminating feature of the Galactic FRB source is that the magnetar is embedded in a SN remnant, providing direct evidence of a core-collapse SN (CCSN) progenitor \citep{Gaensler14, Israel16, Kothes18}. CCSNe are the dominant mechanism for forming neutron stars, some of which can acquire magnetar field strengths of $\gtrsim 10^{14}$~G \citep{DuncanThompson92, Egorov09, Mereghetti20}. Consistent with this, eleven magnetars have confirmed associations with SN remnants observationally (\citealt{Rea25} and references therein). Hence, if some FRBs are powered by young magnetars, then most FRBs should be coincident with a past CCSN. Moreover, the past SN must precede the FRB by a timescale ranging from years to decades, allowing the surrounding plasma to become transparent enough for the radio emission to escape \citep{Zhang2023}.

At present, substantial efforts have been largely dedicated to searches for prompt optical emission, afterglows, or SNe {\it following} FRBs \citep{Marnoch20, Kilpatrick21, Nunez21, Li22, Hiramatsu23, Kilpatrick24}. However, no definitive connection has been established thus far between any FRB and a \textit{preceding} CCSN or optical transient. This notable gap stems from the lack of systematic searches for optical counterparts to FRBs such as past SNe. Leveraging the availability of wide-field synoptic surveys and, soon, the Legacy Survey for Space and Time (LSST; \citealt{LSST}), such searches can be conducted systematically for the first time. In particular, these efforts are capable of discovering SNe that are broadly dichotomized into CCSNe and Type Ia SNe (SNe Ia), traced to either the explosion of a massive star or the thermonuclear runaway of a white dwarf (WD), respectively \citep{Nomoto84, Smartt09, McCully14, Smartt15, Shen18}. Therefore, the robust association of even a single FRB to a past SN would form a definitive link to specific stellar evolutionary pathways and provide some of the most direct evidence for their origins. 

Coupled with the expected large and uncertain temporal offset between past SNe and FRBs, this matching challenge is further compounded by the broad range in localization precision across radio facilities that discover FRBs, spanning milliarcsecond (mas) to arcminute scales (e.g., \citealt{Marcote17_R1, CHIMEcat1, Nimmo22b, law2024, Shannon24}). Currently, well-localized FRB sources make up a small fraction of the total FRB population. The Canadian Hydrogen Intensity Mapping Experiment (CHIME/FRB, \citealt{CHIMEcat1, CHIMEBB}) is the most prolific FRB discovery survey, with a detection rate of $\sim$1000 FRBs per year. More recently, CHIME/FRB has deployed a series of three CHIME-like Outrigger telescopes spread across the North American continent to address the FRB localization limitation \citep{OutriggerOverview}. These outrigger stations are designed to facilitate very long baseline interferometry (VLBI), thereby enabling precise FRB localizations for hundreds of events per year for the first time \citep{OutriggerOverview}. The first catalog of VLBI-localized FRBs has already been produced by the \kkoname~Outrigger (KKO) station\footnote{The name of the first Outrigger, \kkoname{}, was a generous gift from the Upper Similkameen Indian Band and means ``a listening device for outer space.''}  \citep{Lanman24}, advancing the number of well-localized FRBs and enabling precise associations with their host galaxies \citep{KKOhost25}. 

%describe what each section does. 
In this work, we present searches for coincidences in position and redshift between FRBs and past SNe, as well as FRBs and other optical transients. These searches are enabled by the machinery we developed, applied to the first sample of 83 CHIME-KKO FRBs and $\approx 100$ of the well-localized FRBs known to date from the literature. We organize the paper as follows. In Section \ref{sec:data}, we describe our sample of CHIME-KKO and literature FRBs, together with simulated FRBs, and the optical transient catalog. Next, we outline the methodology for identifying optical counterparts that are both positionally and redshift coincident with the FRB sample in Section \ref{sec:crossmatch}. In Section \ref{sec:results}, we present the results of the search using the observed FRB sample and evaluate the probability of chance coincidence based on a simulated FRB population. We assess the timescale sensitivity of the transient catalog relative to the FRB sample and compare their redshift distributions in Section \ref{sec:discuss}. Finally, we summarize our conclusions in Section \ref{sec:conclusions}.

\begin{figure*}[t!]
    \centering
    \includegraphics[width=\textwidth]{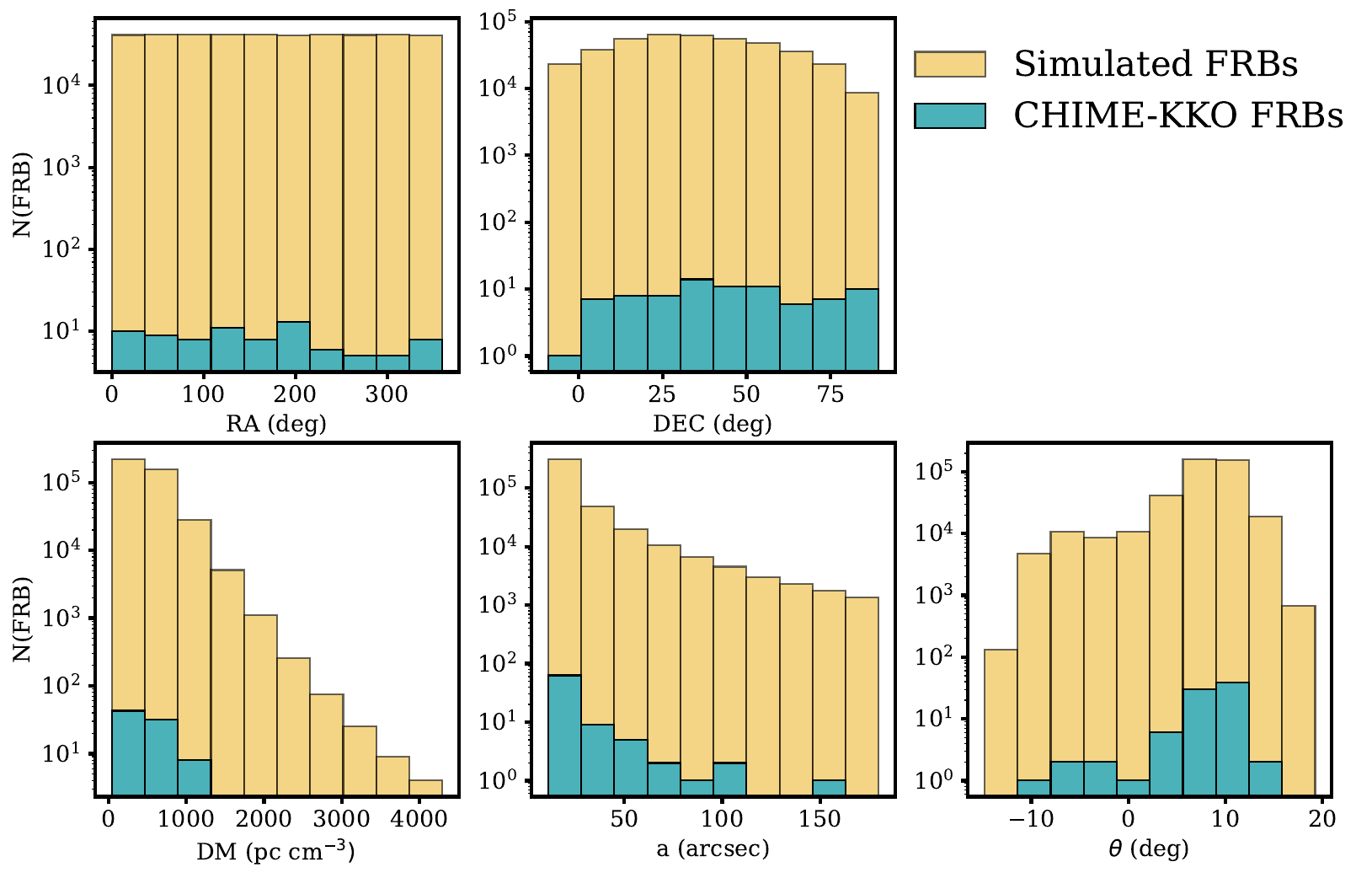}
    \caption{Distributions of the FRB position, DM, semi-major axis ($a$), and orientation angle ($\theta$) of the localization ellipses for a total of 415,000 simulated FRBs (yellow) in comparison to the observed distributions of 83 CHIME-KKO FRBs (blue). 1D KS tests indicate that our simulation accurately models the distribution of these parameters, except for declination, which is more aligned with those of FRBs from the first CHIME/FRB catalog.}
    \label{fig:hist_sim_real}
\end{figure*}

\section{Data} \label{sec:data}

\subsection{Observed FRB Samples} \label{sec:obsfrbsamples}
We use a sample of 83 distinct FRB sources localized by the CHIME-KKO system \citep{KKOhost25} to identify positionally coincident optical transients. These FRBs are from the first observational campaign of the Outriggers program, conducted between December 9, 2023 and February 10, 2024. The $\sim$66 km distance between the KKO station and the primary CHIME site enables VLBI, thus significantly improving the uncertainties in the FRB localization relative to those from the first CHIME/FRB catalog \citep{CHIMEcat1, CHIMEBB}. The KKO Outrigger achieves arcsecond-level accuracy along the baseline axis and a median localization ellipse area of 113 arcsec$^2$. Within our CHIME-KKO FRB sample, 80 are new, as-yet non-repeating FRB sources and three are repeat bursts from known sources: FRBs\,20190303A, 20191106C, and 20220529A (\citealt{collaboration2023chime, michilli2023subarcminute}; Cook et al. \textit{in prep}). 
To ensure a comprehensive sample of well-localized FRBs, we also compile FRBs from the literature with localization uncertainties of $<$~3\arcsec. Beginning with the most precise, mas-localized repeating FRBs, we include FRBs\,20121102A, 20180916B, 20200120E, 20201124A, and 20220912A \citep{Marcote17_R1, Marcote20_180916B, Kirsten22, Nimmo22, Hewitt24_220912A}. Next, we incorporate all other known subarcsecond- and arcsecond-localized FRBs \citep{Bhandari22, Niu22, Bhandari23, Hewitt24_R12, Shah24, Snelder_R147}. Finally, we include all FRBs from the Deep Synoptic Array (DSA; \citealt{Connor24, law2024, Sharma24}) catalog, the Australian Square Kilometer Array Pathfinder (ASKAP; \citealt{Shannon24}) catalog, and the MeerKAT telescope \citep{Driessen24, Rajwade24, Tian24}. The final sample consists of 93 well-localized FRB sources.

\subsection{Simulated FRB Sample} \label{sec:simfrbsamples}

To quantify the likelihood of an FRB-transient match by chance and to build the probability of chance coincidence ($P_{cc}$) curve as detailed in Section \ref{sec:Pcc}, we generate a population of simulated CHIME-KKO FRBs. Following the approach outlined by \cite{Curtin23}, we simulate 5000 sets of 83 FRBs (415,000 total) using the probability distributions in declination ($Dec.$) and time covered by the first CHIME/FRB catalog \citep{CHIMEcat1}. For $Dec.$, we draw uniform random values and transform them using the inverse cumulative distribution function (CDF) of the observed $Dec.$ distribution. For the time distribution, we randomly sample timestamps within the range covered by the first CHIME/FRB catalog. Each simulated FRB is then assigned a right ascension ($R.A.$) based on a random offset from the meridian between $-$1.6$^{\circ}$ and $+$1.6$^{\circ}$ at the simulated time of observation, corresponding to the full width half maximum of the CHIME primary beam at 400~MHz.

The simulated FRBs are characterized by the following parameters: $R.A.$, $Dec.$, dispersion measure (DM), and localization ellipse described by semi-major axis ($a$), semi-minor axis ($b$), and orientation angle ($\theta$) measured east of north. We employ \texttt{scipy.stats} to model the observed CHIME-KKO distributions from \citet{KKOhost25} for each parameter. In particular, we fit the DM and $a$ with log-normal functions, and use Gaussian kernel density estimation with a kernel width of 0.4 degrees for $\theta$ to best characterize their respective distributions. As discussed by \cite{KKOhost25}, the semi-minor axis ($b$) astrometric uncertainty is 2\arcsec~at the 1$\sigma$ level, and therefore we fix $b$ of all simulated FRBs at this value. Figure \ref{fig:hist_sim_real} shows the distribution of each parameter for both the simulated and observed CHIME-KKO FRBs. 

To validate that the two distributions are drawn from the same underlying population, we performed 1D Kolmogorov-Smirnov (KS) tests for each parameter. The null hypothesis assumes that the two samples are drawn from the same distribution, and we reject it if the p-value ($P_{\mathrm{KS}}$) is less than 0.05. The KS tests confirm statistical consistency between the observed and simulated samples for all parameters except $Dec.$, where $P_{\mathrm{KS}} = 0.02$. However, the positions of the simulated FRBs are generated from the first CHIME/FRB catalog, with a declination distribution that follows the sensitivity function of CHIME. Although we have not yet tested for systematic biases in the Outrigger systems, we do not expect strong declination difference between FRBs detected by CHIME and those detected by both CHIME and the KKO Outrigger as the sample grows.

\subsection{Optical Transients Catalog}\label{sec:optcatalog}
The Transient Name Server (TNS)\footnote{\url{https://www.wis-tns.org/}} is a public platform for reporting new optical transients, including SNe, with over 16,000 classified SNe to date \citep{Gal-yam21}. When available, each transient record contains information such as host galaxy identification, discovery date, and observed spectra. 
Its extensive and continuously growing dataset provides an ideal ground for identifying potential FRB optical counterparts. Additionally, we download a separate sample from TNS comprising the past SNe ingested from the Central Bureau for Astronomical Telegrams (CBAT)\footnote{\url{http://www.cbat.eps.harvard.edu/lists/Supernovae.html}}, which are not part of the CSV-formatted staged data, and merge all the objects into a single transient catalog. The CBAT SNe sample includes all events reported from 1885 through 2015.

In Table \ref{tab:steps}, we track the total number of optical transients at each stage, progressively narrowing down the catalog by applying various criteria to exclude unlikely counterpart candidates. As indicated in step 0, the initial sample comprises approximately 152,000 transients from the TNS database. From this pool, we discard objects classified as FRBs, galaxies, variable stars, M-type dwarfs, active galactic nuclei, and other sources not relevant to this work from the catalog. We also excise any objects in the Galactic plane ($|b| < 10^\circ$) where surveys are highly incomplete for extragalactic transients \citep{Fremling20}\footnote{We removed this criterion for observed CHIME-KKO and literature FRBs as some have been observed at low Galactic latitude (e.g., FRB\,20150215A; \citealt{Petroff17})}. In fact, due to MW extinction and the overdensity of stars in this region, most unclassified objects are likely foreground Galactic transients. The final catalog of optical transients (and possibly viable SNe) consists of 141,869 objects.

\begin{deluxetable*}{ccccc}[t!]
\linespread{1.2}
\tablecaption{Analysis matching TNS transients to simulated FRBs at 5$\sigma$ localization region \label{tab:steps}}
\tablecolumns{3}
\tablewidth{0pt}
\tablehead{
\colhead{Step} &
\colhead{Criterion to Satisfy} &
\colhead{Total Candidates} & 
\colhead{Final Candidates} &
\colhead{Rejected Candidates}}
\startdata
\hline
0 & Classified as a relevant transient in TNS and the Galactic latitude $|b|$ is $\ge 10^\circ$ & 152,070 & 141,869 & 10,201 \\ % updated
1 & Within the 5$\sigma$ FRB localization region & 141,869 & 408 & 141,461 \\
2 & Not classified as a minor planet in MPCORB & 408 & 408 & 0 \\
3 & Not classified as a variable star in ASAS-SN or Gaia DR3 catalogs & 408 & 393 & 15 \\
4 & Not classified as an AGN or a quasar in MILLIQUAS or SDSS catalogs & 393 & 378 & 15 \\
5 & Host information is available  & 378 & 369 & 9 \\ 
6 & Robust host association of $P(O|X)$ $\ge 90\%$  & 369 & 281 & 88 \\ 
7 & Redshift is consistent within the 95$\%$ confidence level  & 281 & 254 & 27 \\ 
\enddata

\tablecomments{In step 0, we do not consider the Galactic latitude cut for the observed CHIME-KKO and literature FRBs. The number of transients that we used to calculate the $P_{cc}$ values in Figure \ref{fig:pcc} is slightly smaller than the number of total candidates in steps 4 and 7 since in a few cases, there can be more than one match in a given set of simulated CHIME-KKO FRBs.}

\end{deluxetable*}

\section{Identifying Positional and Redshift Coincidences Between FRBs and Optical Transients}
\label{sec:crossmatch}

\begin{figure}
    \centering
    \includegraphics[width=\linewidth]{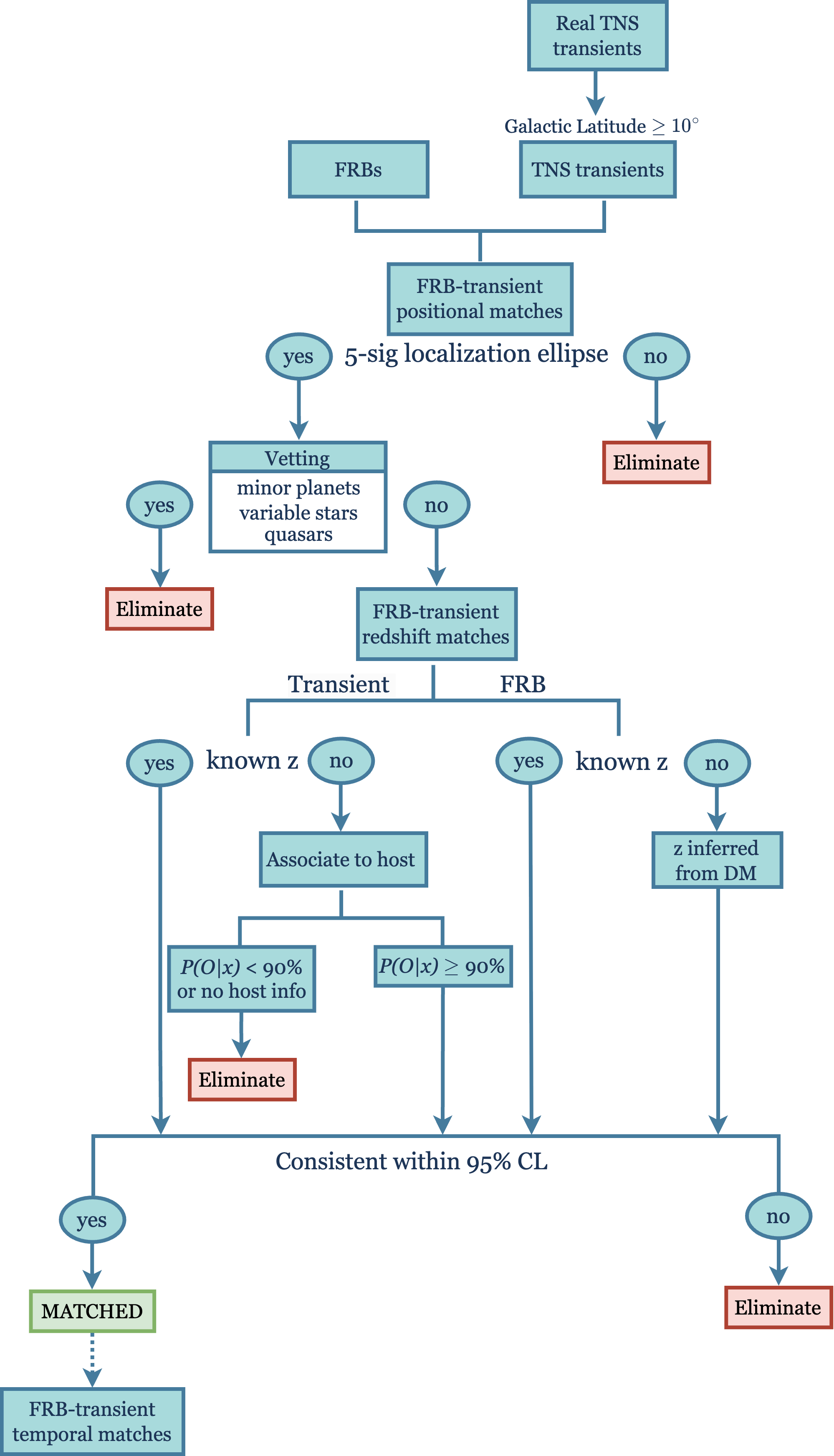}
    \caption{Workflow diagram illustrating the systematic search for positional and redshift coincidences between an FRB and an optical transient. All steps are described in Section \ref{sec:crossmatch}. The number of candidates remaining after each step for the simulated FRB population at the 5$\sigma$ localization significance is provided in Table \ref{tab:steps}.}
    \label{fig:flowchart}
\end{figure}

In this section, we outline the criteria for identifying TNS transients as potential optical counterparts to FRBs, based on the localization region, redshift, and time separation. The total candidate count at each stage of our vetting process is listed in Table~\ref{tab:steps}. The workflow of the pipeline, and the criteria considered at each step, is illustrated in Figure \ref{fig:flowchart}. We address steps for both real and simulated FRBs in the following sections.

\subsection{Positional Search} \label{sec:positional}
To identify matches between a CHIME-KKO FRB and an optical transient, we first check for positional coincidence. In step 1, we perform a cone search with a 15$\arcmin$ radius centered on the FRB position, which encompasses the largest 5$\sigma$ localization ellipse ($a$ $\approx$ 3$\arcmin$) as a first pass. This cone search is executed by querying TNS via its application programming interface (API) to retrieve optical transients. 

Similarly, we perform the cone search for literature FRBs using a 15\arcsec\ search radius set by the largest localization uncertainty (3\arcsec) in the sample. In contrast to the observed CHIME-KKO FRBs, the localization uncertainties of literature FRBs are smaller or comparable to those of the optical transients. In such cases, systematic astrometric uncertainties start to dominate the difference between the FRB and transient positions. These uncertainties arise from the process of aligning optical images to absolute reference frames (e.g., matching stars to their known positions in the {\it Gaia} catalog; typically $<0.3\arcsec$) when determining the transient position. To account for these systematic uncertainties, we impose an FRB positional uncertainty floor of 1\arcsec, as the typical positional accuracy of the transient is $<$1\arcsec\ \citep{ZTF}.

For the simulated FRBs, instead of the API, we take advantage of the daily CSV-formatted staged data provided by TNS, downloading the entire database as of September 17, 2024 for local positional coincidence search (cross-matching). Given the large number of simulated FRBs, this approach mitigates the time delays caused by server throttling when handling a large volume of requests. We choose this date as the cutoff because it includes the most recent FRB discovery in our CHIME-KKO sample. To perform the cone search, we use \texttt{scipy.BallTree} for fast nearest-neighbor queries. This data structure enables rapid positional cross-matching by recursively splitting the 2D ($R.A.$ and $Dec.$) tree along each dimension and eliminating those regions that are beyond the search radius of 15$\arcmin$ (``pruning''). We employ the haversine formula to account for spherical geometry when computing the angular distances between FRBs and optical transients. Objects located outside the cone search region are discarded.

\begin{figure}
    \centering
    \includegraphics[width=\linewidth]{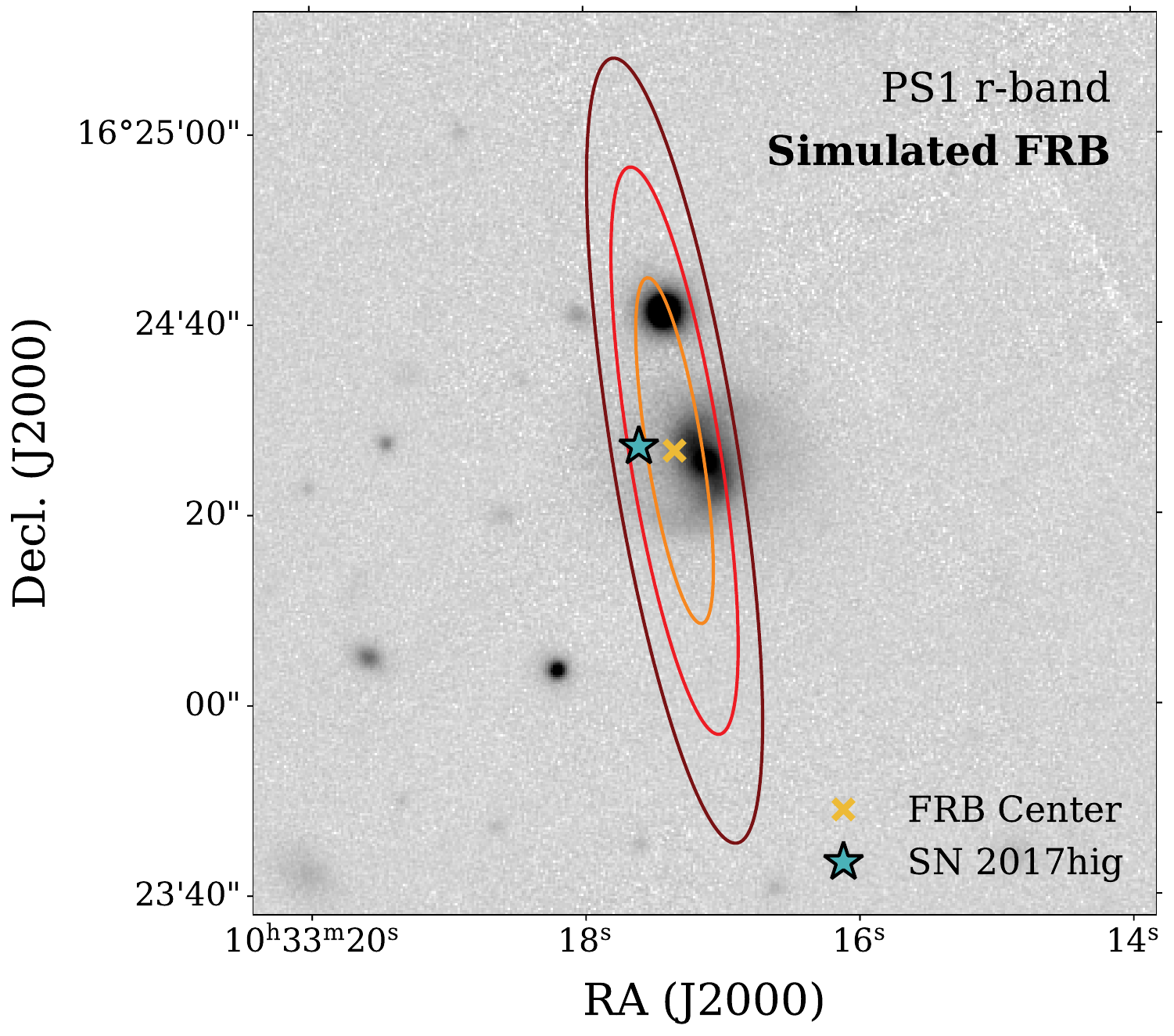}
    \caption{Archival Pan-STARRS1 $r$-band imaging of the host galaxy of SN\,2017hig is shown, with the SN position marked by the blue star. The 1$\sigma$, 2$\sigma$, and 3$\sigma$ localization contours of the simulated FRB are shown in orange, red, and maroon, respectively. The center of the FRB localization is denoted as a yellow cross. In this example, SN\,2017hig is positionally matched within the 2$\sigma$ localization ellipse of the FRB.}
    \label{fig:loc}
\end{figure}

In the final part of step 1, we refine the positional coincidence matching by reducing the search area to the FRB localization ellipses. We require that the transient position overlaps with the 5$\sigma$ localization region of the FRB. The FRB positional uncertainty is characterized by its ellipse parameters ($a$, $b$, and $\theta$). Assuming the localization uncertainties are Gaussian, we compute the covariance matrix of each localization ellipse in order to determine significance levels extending out to 5$\sigma$. We then calculate the Mahalanobis distance, which measures how many standard deviations the transient lies from the FRB center, accounting for the elliptical shape of the FRB localization uncertainties. Any optical transient beyond the 5$\sigma$ localization boundary is eliminated. Figure \ref{fig:loc} shows an example of a simulated FRB and transient positional coincidence. By the end of step 1, we have a remaining catalog of 408 possible transient matches for the simulated FRB population. The code developed in this work to search for positional coincidences between FRBs and optical transients from TNS is publicly available on GitHub \citep{zenodo_link}. %\footnote{\url{https://github.com/FRBs/FRB/tree/main/frb/scripts/SN-FRB}}.

\subsection{Vetting Known Contaminant Transients}
The next step is to remove objects that are unclassified in TNS, but are classified in other catalogs as minor planets, variable stars, or AGN, thus precluding their association to FRBs (step 2 in Table~\ref{tab:steps}). Step 0 in Table \ref{tab:steps} removes a portion of classified objects not relevant to this work, and around 75$\%$ of TNS objects are in this category. 
In an effort to improve the purity of our optical transient catalog, we cross-match the remaining matched candidates after step 1 with the relevant public catalogs. This vetting process corresponds to steps 2-4 in Table \ref{tab:steps}.

For near-Earth minor planets in step 2, we query the IAU Minor Planet Center Orbit Database (MPCORB)\footnote{\url{http://www.minorplanetcenter.net/iau/MPCORB.html}} within a 30$\arcsec$ radius of the transient position on the day each transient is discovered. This search radius is chosen as it represents the standard criterion for vetting minor planets among extragalactic transients \citep{Rastinejad22}. In step 3, we query variable stars from the \textit{Gaia} Data Release 3 (DR3; \citealt{GaiaDR3}) along with the All-Sky Automated Survey for Supernovae (ASAS-SN; \citealt{Jayasinghe18, Jayasinghe19}) catalogs at the location of each transient using a 3$\arcsec$~radius. For \textit{Gaia} DR3 objects, we classify a TNS transient as stellar if it satisfies any of the following conditions: (i) flagged as variable, (ii) total absolute proper motion $>$ 3 $\times$ proper motion error, and (iii) parallax significance $>$ 8 (see \citealt{Tachibana18} for more details).

In step 4, we query the Million Quasar catalog (MILLIQUAS; \citealt{Flesch15, Flesch21}) and the Sloan Digital Sky Survey Data Release 16 (SDSS DR16; \citealt{Lyke20}) at the location of every transient using a search radius of 3$\arcsec$ for AGN and quasars, which we do not consider likely FRB counterparts in this context (e.g., \citealt{Vieyro17, Zhang18}). Specifically, we exclude TNS transients if they are labeled as quasi-stellar objects in SDSS DR16 or their probability of being a quasar exceeds $>97\%$ in MILLIQUAS, as this threshold has been shown to correlate well with confirmed quasars in SDSS \citep{Flesch15}. Out of the 408 transients, 30 are cross-matched as either variable stars or AGN/quasars, while none are identified as a minor planet. This means that 378 transients make it through step 4.

\subsection{Optical Transient and Host Associations} \label{sec:hostassoc}
At step 5 of our FRB-transient association pipeline, we establish positional coincidence for a set of FRB-transient matches and vet them based on known catalogs. To determine whether the candidate match is plausible, we incorporate host galaxy redshift for all optical transients in the catalog to check whether the redshift of the transient is consistent with the redshift or redshift estimate of the FRB. This section describes steps 5 and 6 in Table~\ref{tab:steps}.

For transients in the TNS catalog that have passed step 4, we check if a redshift is available. Classified SN spectra often include redshift information, which is reported to TNS. Moreover, roughly half of the past SNe from CBAT have a reported host galaxy. However, the redshifts of these hosts are not part of the catalog.  
If the transient has no direct redshift information, then we assign each transient to its most probable host using the Probabilistic Association of Transients to Their Hosts framework (PATH; \citealt{Aggarwal2021}). PATH is a Bayesian method for transient-host association that relies on an ensemble of observable features, including sky positions, angular sizes, and galaxy brightness. For uniformity in our transient catalog, we also apply PATH to the CBAT SNe.

To run PATH, we utilize public, deep, and wide-field surveys to retrieve images of the transient field. We first query the Dark Energy Camera Legacy Survey Data Release 8 (DECaLS DR8; \citealt{Decals}) in \textit{r}-, \textit{g}-, and \textit{z}-band because of its superior depth ($m_r = 23.5$ mag). If no DECaLS images are available, we default to the shallower ($m_r = 23.2$ mag) \textit{r}-band images from Pan-STARRS Data Release 2 (PS1 DR2; \citealt{Chamber16, PS1}), as all CHIME and simulated FRBs lie within its footprint.

In step 5, we extract a 1$\arcmin$ $\times$ 1$\arcmin$ region around the transient, assuming a conservative transient localization uncertainty of 3$\arcsec$ for galaxy candidate selection since the typical positional accuracy of the transient is $\approx$~1\arcsec~\citep{ZTF}. 
Since PATH does not inherently distinguish between point-like and extended sources, to minimize mistaking stars for galaxies, we apply multiple criteria to reject stellar sources. For DECaLS images, we avoid bright foreground stars by selecting sources fainter than 11 mag. Additionally, using flagging labels from \textit{Gaia} DR3, we remove already-identified point sources. If any object has a morphological type of point spread function (PSF) in DECaLS, we similarly exclude it. We apply three additional criteria for PS1 images. First, we validate that each source has a \cite{Kron80} magnitude and radius, and compare the \texttt{Kron} magnitude to the PSF magnitude. An object is considered stellar if $m_\textrm{PSF} - m_\textrm{Kron} < 0.05$~mag \citep{Farrow14}. Second, we rely on the PSF likelihood metric, excluding an object if the log-transformed PSF likelihood satisfies $\log(\mathcal{L}_{\textrm{PSF}}) > -2$. Lastly, we query the PS1 point source catalog (PS1-PSC; \citealt{Tachibana18}) that contains $\sim$1.5 billion sources from the PS1 first data release, along with their classification as either extended ($P=0$) or point ($P=1$) sources. We discard any source with a point-source probability ranking $P>0.20$.

After removing all possible stellar objects, we assign a probabilistic association of nearby galaxies for each transient as part of step 6. The PATH formalism applies Bayes' rule to calculate posterior probabilities, $P(O|x)$, for each candidate galaxy $O$. Based on existing FRB host associations using PATH \citep{law2024, craft_ics_paper}, we apply an ``inverse'' prior to accommodate the high density of faint galaxies and an ``exponential'' angular offset prior, scaled by $1/2$ of the effective radius and truncated at six effective radii of candidate galaxies. We assume the probability that the host is unseen, $P(U)=0.1$ for DECaLS and 0.2 for PS1 images, reflecting their respective survey depths. For each transient, we identify the galaxy with the highest $P(O|x)$, and only rely on host associations when they have relatively high probabilities, $P(O|x) \geq 90\%$ \citep{Aggarwal2021}.

We note that the most probable hosts for the past SNe are consistent with those listed in CBAT. However, the PATH posterior probabilities for the hosts of SN\,1995ai and SN\,1987M fall below our threshold of $P(O|x) = 90\%$. Therefore, we excluded both events from our cross-matching. Out of 393 positional matches to the simulated FRBs, a large majority (369 transients, or $\sim 94\%$) have host and/or redshift information. The redshifts of the positionally matched past SNe span from $z = 0.004$ to $z = 0.663$.

\subsection{FRB Redshift Search} \label{sec:redshift_frb}
After completing the transient-host associations in steps 5 and 6, we extract redshift information for both the FRB and the optical transient to complete the final step in Table \ref{tab:steps}. In this section, we focus on the redshifts of FRB samples. For FRBs (both simulated and real) that are positionally coincident with TNS transients, if the FRB has a robust host association, we adopt the corresponding redshift measurement directly. However, to overcome the challenge of FRBs lacking host associations or redshifts, particularly for our simulated FRB population, we apply the Macquart relation. This relation correlates the FRB's DM from the cosmic web with the redshift of its host galaxy \citep{Macquart20}. Despite the apparent scatter in the Macquart relation \citep{Baptista24}, DM remains a useful proxy for cosmological distances, especially for well-localized FRBs in the low-redshift regime.

We now describe the method used to infer the redshift of the FRB from its DM. The FRB DM can be decomposed as DM$_\mathrm{FRB}$ = DM$_\mathrm{ISM}$ + DM$_\mathrm{Halo}$ + DM$_\mathrm{EG}$. The DM contributions from the MW's interstellar medium and halo are captured by DM$_\mathrm{ISM}$ and DM$_\mathrm{Halo}$, while DM$_\mathrm{EG}$ represents the extragalactic contribution, including that from the intergalactic medium and the host galaxy. We use the NE2001 model of \cite{NE2001} to estimate the disk contribution (DM$_\mathrm{ISM}$). The DM contribution from the MW halo is highly uncertain, ranging over 10 -- 111\,pc~cm$^{-3}$ \citep{Dolag15, Prochaska19, Keating20, Cook23}. Here, we assume a typical value of DM$_\mathrm{Halo}$ = 30\,pc\,cm$^{-3}$ \citep{Dolag15}. Taking into account these contributions to the total DM of the FRB, we estimate a joint probability distribution of the redshift and DM$_\mathrm{EG}$ following the framework developed by \cite{james2023_chime_zDM}. 

\begin{figure*}
    \centering
    \includegraphics[width=\linewidth]{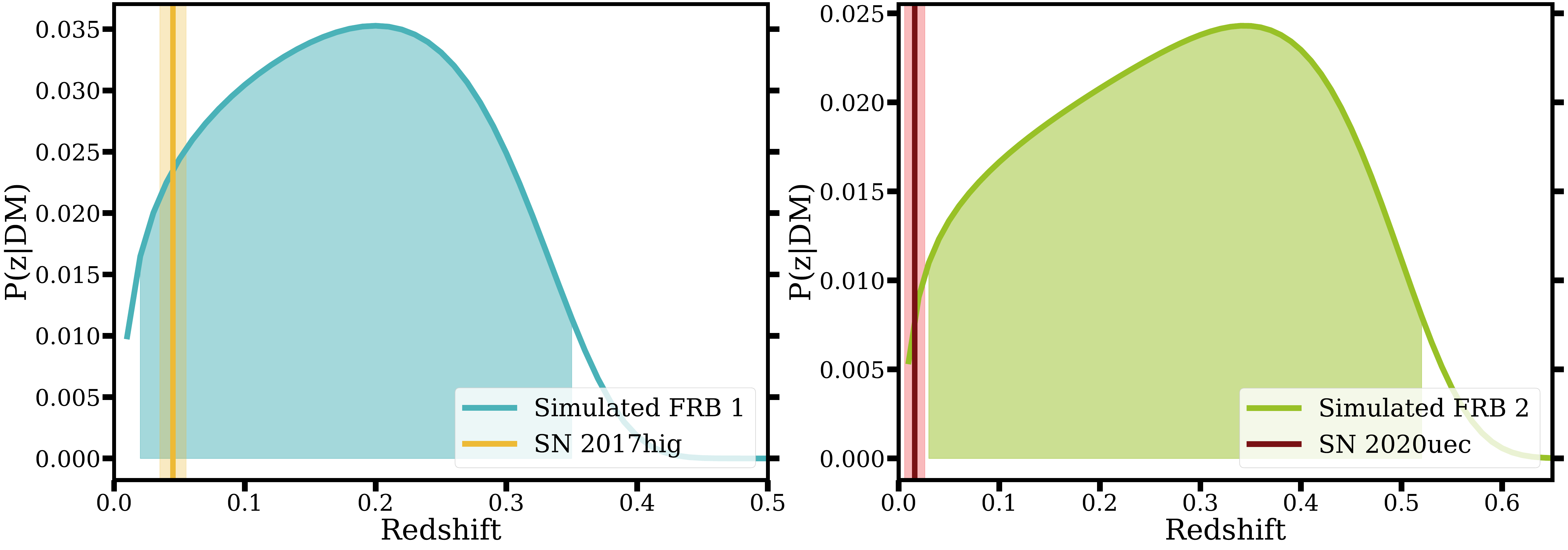}
    \caption{\textit{Left}: Probability distributions of the redshift inferred from the total DM of a simulated FRB using the framework developed by \cite{james2023_chime_zDM}. The shaded blue region represent the 95$\%$ confidence interval. The spectroscopic redshift of SN\,2017hig is shown as a solid yellow line, with the shaded area denoting the assumed redshift uncertainty of 0.01.
    The overlap between the transient redshift and the DM-inferred FRB redshift distribution indicates redshift consistency. This demonstrates that SN\,2017hig satisfies all criteria outlined in Table \ref{tab:steps}, and a positional and redshift coincidence is established for simulated FRB 1. \textit{Right}: Same as left, but for a different simulated FRB positionally coincident with SN\,2020uec. The red line marks the spectroscopic redshift of the SN along with its associated uncertainty. In contrast to the left panel, the transient redshift does not fall within the 95$\%$ confidence level of the FRB redshift distribution, and thus the transient is not redshift coincident with this simulated FRB.}
    \label{fig:pzDM}
\end{figure*}

\subsection{Transient Redshift Search}\label{sec:redshift_transient}
We turn to the remaining work to complete step 7, which involves obtaining redshift information for transients that passed step 6. As mentioned in Section \ref{sec:hostassoc}, classified SNe have spectroscopic redshift ($\mathrm{z_{spec}}$) measurements reported in TNS. For these values, we adopt a standard uncertainty of 0.01 \citep{Blondin07}. Otherwise, we search for information on $\mathrm{z_{spec}}$ and its uncertainty for observations within a 3$\arcsec$ radius of the most probable host for each transient. We acquire all $\mathrm{z_{spec}}$ values from the NASA/IPAC Extragalactic Database (NED)\footnote{\url{https://ned.ipac.caltech.edu/}}.

For galaxies without $\mathrm{z_{spec}}$, we leverage available $\mathrm{z_{phot}}$ estimates from the PS1 Source Types and Redshifts with Machine learning (PS1-STRM; \citealt{Beck21}) catalog. The main advantage of PS1-STRM is its use of broad-band $grizy$ filters and its coverage of three-quarters of the sky, encompassing over three billion sources. We ensure the reliability of the $\mathrm{z_{phot}}$ estimates by only retaining galaxies with a relative uncertainty of $\mathrm{z_{photErr}}$/$\mathrm{z_{phot}}<1$. If both NED and PS1-STRM yield null values, we make a final attempt to extract $\mathrm{z_{phot}}$ from the Dark Energy Spectroscopic Instrument (DESI) Legacy Imaging Surveys Data Release 8 \citep{Duncan22}.

After obtaining the redshifts for both the FRB (Section \ref{sec:redshift_frb}) and the transient, we assess whether the redshifts are consistent. In step 7, we define a redshift coincidence when the transient redshift, within its 1$\sigma$ uncertainty range, overlaps with either (i) the FRB redshift (as measured from the host) or (ii) the DM-inferred redshift range (within the 95\% confidence interval). We recognize that this comparison is not statistically symmetric. However, we choose a broader range for the FRB DM-inferred redshift distributions because the associated uncertainties are substantially larger than those of the optical transients. We illustrate this overlap in redshift between an FRB and a transient in Figure \ref{fig:pzDM}. Specifically, we show two examples of simulated FRBs that are positionally coincident with a SN in which one also matches in redshift (SN\,2017hig), while the other does not (SN\,2020uec). Based on this criterion, we eliminate 27 matches, resulting in a final sample of 281 FRB–transient matches for the simulated FRB population. The remaining sources are either highly offset from viable galaxy candidates or their most probable hosts are too faint for a robust association.

\subsection{Temporal Search} \label{sec:time}

Finally, for the purpose of identifying past SNe that may be progenitors of the FRBs in this work, we search for transients that occurred before each FRB. To quantify the temporal separation, we use the discovery dates of every optical transient and FRB that are a positional and redshift match. For optical transients, we obtain discovery dates from TNS as part of the metadata extraction in Section \ref{sec:optcatalog}. CHIME-KKO and literature FRB sources have well-defined arrival times due to their extremely short durations, with detection dates represented in their names following the TNS convention. In contrast, assigning realistic discovery dates, which represent only an upper limit on when the progenitor began emitting FRBs, to our simulated FRB population is not feasible, as they are meant to represent a much longer time baseline than the operational period of CHIME-KKO Outrigger. Therefore, we did not perform this temporal check for our simulated FRBs. 

Although our infrastructure is designed to search for matches between past SNe and FRBs, it can be readily customized to accommodate any temporal bounds, offering the flexibility to explore a wide range of progenitor models that predict optical counterparts on different timescales. For example, this work can be implemented to search for a luminous, red nova or fast blue optical transient predicted by the hyper-accreting X-ray binary model months to years following repeating FRB sources \citep{Sridhar21, Sridhar22}. It can also be used to search for a bright optical flash lasting $\lesssim$ 1s, occurring contemporaneously with the FRB as predicted in the hyperactive magnetar flare scenarios \citep{Beloborodov20}.

\section{Results} \label{sec:results}

\subsection{Transient Searches for Observed FRB Samples}\label{realsearches}
Equipped with the cross-matching tools introduced in this work, we now search for optical transients coincident with CHIME-KKO and literature FRBs. Among the 176 observed FRBs (of which 83 are from CHIME-KKO), our initial cone search (Section \ref{sec:positional}) revealed one significant positional match with a TNS transient. For the subset of CHIME-KKO FRBs specifically, the closest potential match, an unclassified transient AT\,2017ios, lies 1.5\arcmin~away from FRB\,20231101A, corresponding to a 9$\sigma$ offset from the FRB localization ellipse center\footnote{We note that since our normal cut is $5\sigma$ this would not ordinarily be considered a match in our pipeline}. Given this large separation, a physical association is unlikely. Thus, we find no significant associations with the existing CHIME-KKO FRB sample. For literature FRBs, we find a single transient match, AT\,2020hur, which is both positionally coincident and consistent in redshift with FRB\,20180916B. Since no other CHIME-KKO or literature FRBs were found to be positionally coincident with any TNS transients, we did not assess redshift or temporal matches for these sources.

% AT2020hur
The potential match between AT\,2020hur and FRB\,20180916B is inconsistent with FRB magnetar progenitor models linked to past SNe, as the transient occurred 19 months after the first detection of this repeating FRB. However, under alternative progenitor models that predict an optical counterpart following the FRB (e.g., \citealt{Sridhar21}) and in light of a previously reported possible association \citep{Li22}, we consider the match as potentially real. The discovery of AT\,2020hur was first reported on April 8, 2020 by the MASTER-Kislovodsk telescope \citep{Lipunov20} with a Vega magnitude of 18.4~mag and remains unclassified in TNS. Notably, its discovery date coincides with the 6.1-day chromatic active window of FRB\,20180916B, during which the source is expected to be detectable \citep{Pastor21}, thus supporting a possible association with FRB\,20180916B given its spatial coincidence (although the optical transient uncertainty of $\sim 1\arcsec$ is far larger than the FRB positional uncertainty; \citealt{Li22}). However, despite this temporal overlap, the MASTER-Kislovodsk discovery observations of AT\,2022hur do not coincide with CHIME observations of the FRB source, precluding a prompt optical emission temporally coincident with the FRB. 

As an additional check, we examined archival \textit{Hubble Space Telescope} (\textit{HST}) imaging of FRB\,20180916B \citep{Mannings21} and found no apparent point source at the transient position with a limiting magnitude of F110W $>$ 27.3~mag (AB, not extinction-corrected), thereby ruling out the possibility that AT\,2020hur is associated with a foreground star or any other persistent but unassociated source. Given the lack of additional information about the nature of AT\,2020hur, we cannot comment further on the possible association between this source and FRB\,20180916B and consider a detailed analysis of this event to be beyond the scope of this work.

\subsection{Probability of Chance Coincidences}\label{sec:Pcc}

Our search with the initial CHIME-KKO sample and existing literature FRBs did not yield any robust matches to known optical transients. However, if most FRBs are associated with past SNe, this association rate is anticipated to change with the upcoming full deployment of CHIME/FRB Outriggers and the start of the Vera C. Rubin Observatory era. In particular, we expect the detection rates of both well-localized FRBs and optical transients to increase by at least an order of magnitude over the next few years \citep{SKA, LSST, OutriggerOverview}.

As these populations grow, the likelihood of a random match between an FRB and optical transient also grows. It is therefore critical to establish a baseline probability of random coincidences to quantify the expected number of these chance associations. To this end, we construct a $P_{cc}$ curve as a function of the CHIME-KKO localization uncertainty regions using our simulated FRB population. Following the methodology outlined in Section~\ref{sec:crossmatch}, we search for positional and redshift coincidences for all simulated FRBs within localization uncertainty regions corresponding to significance levels from 0.1$\sigma$ to 5$\sigma$ and in increments of 0.1$\sigma$. The $P_{cc}$ values represent the fraction of simulated FRB sets (out of a total of 5000 sets of 83 FRBs) containing at least one transient match. We then determine the uncertainty at each confidence level as the standard deviation in the fractional matches across all sets.

\begin{figure}
    \centering
    \includegraphics[width=\linewidth]{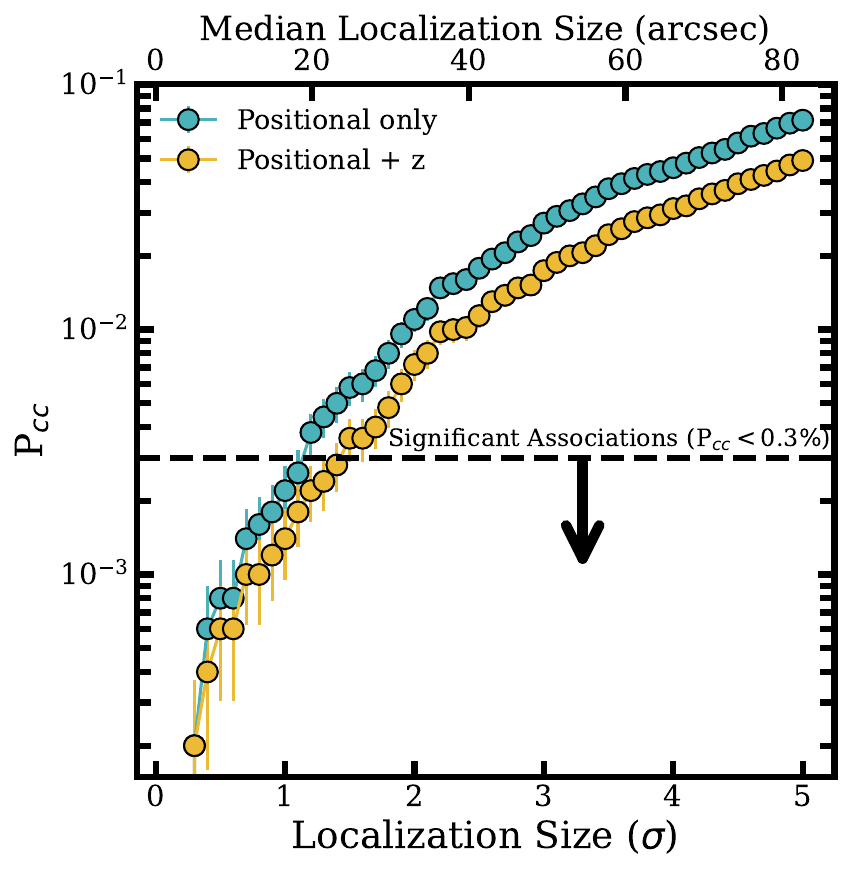}
    \caption{Probability of chance coincidence ($P_{cc}$) curve for positional matches in blue and for combined positional and redshift matches in yellow. The $P_{cc}$ values are a function of the CHIME-KKO localization ellipse sizes in significance levels and represent the fraction of simulated FRB sets with at least one match. The black dashed line represents the false alarm rate of 0.3$\%~P_{cc}$. The error bars represent the standard deviation in the match fractions and are smaller than the symbols beyond $\sim$2.5$\sigma$ (41.3\arcsec). Overall, the $P_{cc}$ values are below 0.1. Any real match within the 1--1.5$\sigma$ (16.5\arcsec -- 24.8\arcsec) localization ellipse would be statistically significant, beyond which the probability of a true association diminishes, and matches become increasingly consistent with random coincidence.} %associations become statistically uninteresting.}
    \label{fig:pcc}
\end{figure}

In Figure~\ref{fig:pcc}, we show the $P_{cc}$ curves for positional coincidences (blue) and for combined positional and redshift coincidences (yellow) as a function of localization ellipse size in units of $\sigma$ and arcsecond for the CHIME-KKO sample. The arcsecond scale is derived by multiplying each $\sigma$ value by the median $a$ of all simulated FRBs. As the FRB localization size (and area) increases, the likelihood of random matches and thus the $P_{cc}$ also increases. We find that the probability of a chance association is generally low ($P_{cc}$ $<$ 0.1), with a maximum value of 0.07 for positional coincidences at a 5$\sigma$ localization size. When incorporating redshift constraints, the entire $P_{cc}$ curve systematically shifts to lower values, with a maximum value of 0.05, as the requirement that each transient has an associated redshift and/or host detection and the redshift itself further limit the number of possible transient matches and search volume of potential matches. Within the 1$\sigma$ (16.5\arcsec) localization ellipse, the $P_{cc}$ values remain  $<$0.001. We consider any association for an individual FRB to be meaningful if it achieves a fiducial false alarm rate of 0.3$\%~P_{cc}$, corresponding to a 3$\sigma$ probability (99.7\%) as represented by the horizontal dashed line. Above this threshold, true matches become indistinguishable from false associations given the localization size. As a result, we find that a transient association is statistically significant only within the 1.1$\sigma$ (18.2\arcsec) localization ellipse for purely positional coincidences and the 1.4$\sigma$ (23.1\arcsec) localization ellipse for positional and redshift coincidences of the CHIME-KKO FRBs.

% how many CHIME-KKO FRBs would need to be discovered before you would expect to see a chance coincidence with an optical transient
The projected FRB detection rate of CHIME/FRB Outriggers is 1--2 sources per day \citep{Lanman24}, implying that once the system is fully operational, it would take only $\sim 1$~month to reconstruct the entire CHIME-KKO sample analyzed in this work. Here, we estimate the number of full-array Outrigger FRB detections required to expect one chance coincidence ($N_{\mathrm{FRB}}^{\mathrm{chance}}$). This can be calculated as 1/$P_{cc}$ described above, which varies based on the localization size considered. 

For associations with SNe, we assume a conservative 1$\arcsec$~localization radius for the FRB. As discussed in Section~\ref{sec:positional}, we adopt 1$\arcsec$ as the lower limit for the combined systematic and statistical uncertainties in FRB localizations and associations with optical transients, even though the localization accuracy is expected to be $\sim$50 mas for most Outrigger-detected FRBs \citep{OutriggerOverview}. This localization uncertainty corresponds to $\sigma = 0.167$ given that the ellipse area $\pi ab \propto \sigma^2$. Since the $P_{cc}$ values scale with the area of the localization ellipses (roughly as $a^{2}$), we employ \texttt{curve$\_$fit} from the \texttt{scipy.optimize} package \citep{Scipy} and fit a power law of the form $\sigma \propto (P_{cc})^{\alpha}$ to the combined positional and redshift $P_{cc}$ curve in Figure \ref{fig:pcc} where $\alpha$ is the power index. This fit allows us to determine the corresponding $P_{cc}$ value for a given $\sigma$. We find a best-fit $\alpha=2.07 \pm$ 0.02. Extrapolating the yellow curve in Figure~\ref{fig:pcc} to $\sigma=0.167$, we derive a $P_{cc}$ value of 4.4 $\times 10^{-5}$. This implies that roughly 1 out of every $\sim 22,700$~FRBs is expected to yield a match with a past SN that is purely a chance coincidence, assuming an FRB localization size of 1\arcsec. 

\begin{figure*}[!t]
    \centering
    \includegraphics[width=\linewidth]{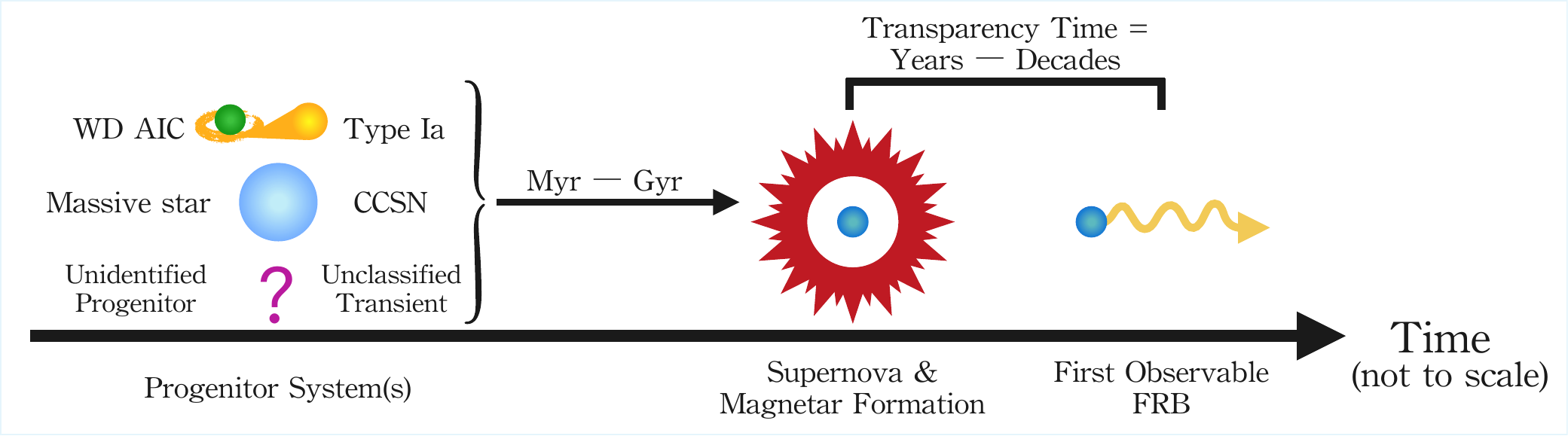}
    \caption{Schematic illustration of possible progenitor systems and transparency times relevant to transient-FRB associations. Various progenitor scenarios are shown on the left, including accretion-induced collapse (AIC) of WDs and massive stars, which lead to different types of SNe and unclassified transients on Myr to Gyr timescales. At the final stage of progenitor evolution, the timeline then spans from SN and magnetar formation to the first observable FRB, highlighting the transparency time, ranging from years to decades, after which the FRB escapes through the surrounding dense plasma environment.}
    \label{fig:timescale}
\end{figure*}

% How long would that take? 
Next, it is useful to estimate how long this might take to accrue such a sample. Depending on the radio instrument, the timescale to find a single chance coincidence match can vary based on its FRB detection rate. In general, this time interval is given by:

\begin{equation} \label{Pcc_equation}
    \Delta t_{\mathrm{FRB}} = \frac{N_{\mathrm{FRB}}^{\mathrm{chance}}}{R_{\text{FRB}}} 
    = \frac{\sigma^{-\alpha}}{A\times R_{\text{FRB}}}
\end{equation}

\noindent where $A$ is the power-law normalization factor and $R_{\text{FRB}}$ is the FRB detection rate. As shown by Equation~\ref{Pcc_equation}, this time interval also depends on the localization size, characterized by the localization uncertainty $\sigma$, and the SN detection rate, which is parameterized by the fitted power law in this work. Assuming the projected CHIME/FRB Outriggers detection rate (1--2 FRB/day), this corresponds to approximately one chance coincidence every 30--60 years for a 1\arcsec~FRB localization uncertainty. The expected number to the first chance coincidence can also be expressed independently of the FRB detection rate as $\sigma^{-\alpha}/A$, which may useful for estimating coincidence rates without converting to a timescale.

At face value, as the number of observed CHIME-Outrigger FRBs increases, the likelihood of a chance association with an unrelated transient also rises. As a result, the anticipated $\Delta t_{\mathrm{FRB}}$ should decrease with the growing FRB detection rate. Indeed, upcoming radio survey telescopes like the Deep Synoptic Array (DSA) 2000 \citep{DSA-2000}, Canadian Hydrogen Observatory and Radio-transient Detector (CHORD; \citealt{CHORD}), and the Square Kilometre Array (SKA; \citealt{SKA, SKA_mac}) are expected to detect and localize $\gtrsim$ 10$^4$ FRBs per year. If these events are also localized to sub-arcsecond precision and this high event rate holds, the expected time interval for a chance coincidence shortens to 1--2 years. However, we note that the precise values of $A$ and $\alpha$ may change based on the increased SN detection rates (for example, once LSST begins), thus changing the time interval. Notably, Equation~\ref{Pcc_equation} can be tailored to any FRB and SN detection rates, as well as FRB localization sizes from other instruments to estimate the expected timescale to achieve at least one chance coincidence.

\section{Discussion}\label{sec:discuss}

\subsection{Supernova Ejecta Transparency Timescale} \label{timscales}

The main goal of our work is to assess the possible connection between past SNe and FRBs. Prior to the emission of FRBs, SNe are expected to lead to the formation of most magnetars, which in turn are expected to produce FRBs. A major uncertainty in this channel is the duration of the delay between the SN explosion and its first observable FRB (hereafter, we refer to this as the ``transparency time''). As illustrated in Figure~\ref{fig:timescale}, following the SN and subsequent magnetar formation, the transparency time spans years to decades \citep{Zhang2023}. During this period, any FRB emission undergoes free-free absorption in the ionized SN remnant, leading to radio scattering and the attenuation of the observable signal. We define the transparency time as the time at which the optical depth drops below unity ($\tau_{\mathrm{ff}} < 1$) to assess the detectability of FRB sources embedded in SNe.

In order to determine when FRB emission can escape the SN ejecta, we consider the free-free transparency timescale ($t_{\mathrm{ff}}$), which is primarily governed by the mass and velocity of the SN ejecta \citep{Murase16, Piro16}. This timescale can also be affected by reionization processes in the SN ejecta, driven by both the reverse shock and internal heating from the magnetar that will increase the reservoir of available scattering electrons \citep{Metzger17}. Assuming a core-collapse SN with an oxygen-dominated ejecta shell (ionized fraction $f_{\mathrm{ion}}$ = 0.4), ejecta temperature of $T_\mathrm{ej} = 10^4$ K, and Gaunt factor of $\bar{g}_{\mathrm {ff}} = 1$, $t_{\mathrm{ff}}$ is given by:

\begin{equation} \label{thinning}
\begin{split}
    t_{\nu}^{\mathrm{ff}} > 6~{\rm yr}~ (1+z)^{3/5}
    \left( \frac{\nu}{600~\rm{MHz}} \right)^{-2/5}
    \left( \frac{M_{\mathrm{ej}}}{0.6\,M_{\odot}} \right)^{2/5} \\
    \times
    \left( \frac{v_{\mathrm{ej}}}{10^4\,\mathrm{km\,s^{-1}}} \right)^{-1} 
\end{split}
\end{equation}\label{eqn:ff}

\noindent where $\nu$ is the emitted frequency, $M_\mathrm{ej}$ is the ejecta mass, and $v_\mathrm{ej}$ is the ejecta velocity in the rest frame \citep{Metzger17}. Crucially, this timescale is expected to be a few to $\sim$tens of years (e.g., observable on human timescales) based on characteristic values of these parameters from observations of SNe with ejecta masses of $\sim$~2$\mathrm{M}_{\odot}$ and ejecta velocities around 5000 km~s$^{-1}$ \citep[as in, e.g.,][]{Gutierrez17}. However, some sub-types of SNe may exhibit longer timescales, as in the case of SN\,1987J \citep{Bietenholz17}.

\begin{figure}
    \centering
    \includegraphics[width=\linewidth]{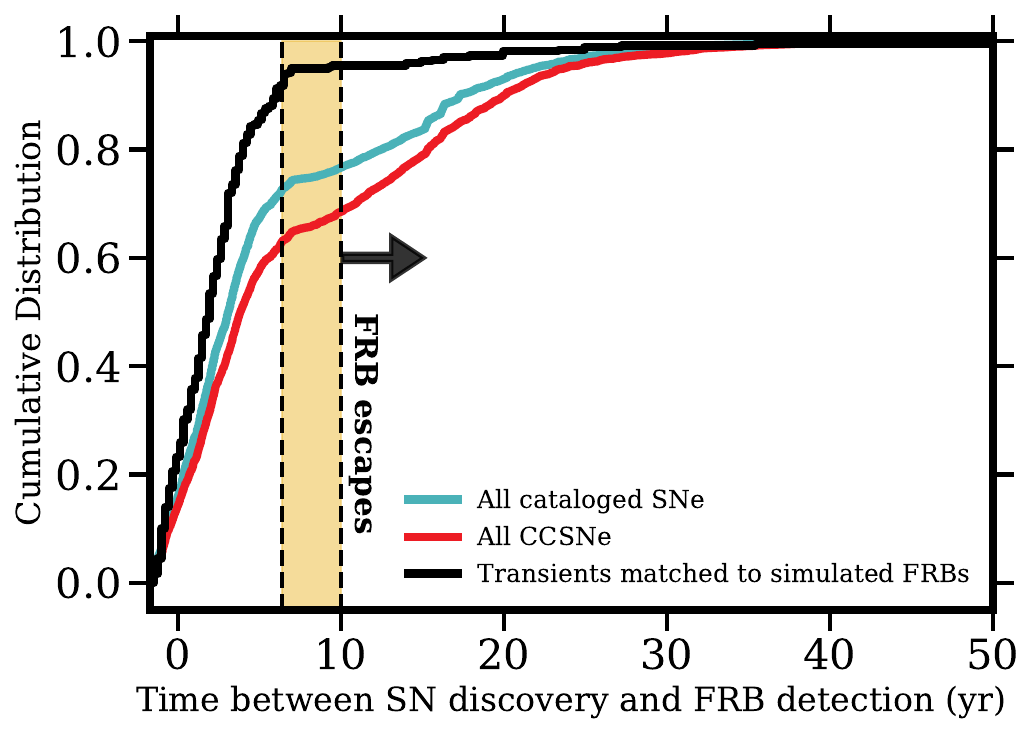}
    \caption{CDFs of the time since SN discovery relative to the simulated FRBs at the reference epoch (January 1, 2023) for positionally-coincident optical transients (black), all cataloged SNe (blue), and CCSNe (red). The yellow shaded region marks a range of reasonable minimum transparency timescale between 6.4 and 10 years, after which the FRB emission could escape the SN ejecta. For transients that are matched to simulated FRBs, $\approx 5-7\%$ of them are older than this transparency timescale and could therefore be viable, astrophysical coincidences. In comparison, $\approx 23-30\%$ of all cataloged SNe and $\approx 32-40\%$ of all CCSNe were discovered earlier than this threshold and could in principle be associated with an observable FRB today. } 
    \label{fig:dd_cdf}
\end{figure}

To explore the transparency timescale on which we are sensitive to SNe preceding FRBs, we next examine the TNS optical transients with respect to the CHIME-KKO FRB sample. As detailed in Section \ref{sec:time}, the simulated FRB population has no discovery dates. To provide a reference epoch to quantify the temporal separations of optical transients and simulated FRBs, we designate January 1, 2023 as the discovery date of all simulated FRBs which marks the beginning of the year in which the CHIME-KKO sample was discovered. In Figure~\ref{fig:dd_cdf}, we plot the CDF of the relative time between the discovery of the FRB and the optical transient. In particular, we show this for all classified TNS SNe, as well as optical transients that are positionally coincident with the simulated FRB population within 5$\sigma$ localization ellipses. We additionally isolate known CCSNe as these are the transients most likely to be linked to FRB sources, and plot their CDF. For clarity, we truncate them at 50~yr, which captures $>$99$\%$ of the TNS SNe and matched optical transient distributions.

As shown in Figure~\ref{fig:dd_cdf}, the times since SN discovery range from $-$1.7 to $>$50~yr, where negative values represent transients with a discovery date after the simulated FRB. We find that optical transients coincident with simulated FRBs have a median time difference of 1.8~yr, whereas the median for all known SNe and CCSNe are 3.1~yr and 3.8~yr, respectively. To place these timescales in the context of ejecta transparency, we define a \textit{minimum} transparency time ($t_{\nu}^{\rm ff}$)
as indicated by the shaded region in Figure~\ref{fig:dd_cdf}. To minimize $t_{\nu}^{\rm ff}$, favorable conditions include low ejecta masses and high ejecta velocities. We use Equation~\ref{thinning} and calculate $t_{\nu}^{\rm ff}$ across a range of ejecta masses and velocities. These parameters are representative of strongly-stripped CCSNe from \cite{Das23}, which have $M_\mathrm{ej} < 1\,\mathrm{M}_\odot$ and likely originate from low-mass helium stars in close binaries. At 600~MHz (CHIME central frequency), we derive a lower limit of $t_{\nu}^{\rm ff}\approx  6.4$ years from SN~2019jak (an example of an SN with favorable parameters to generate a minimum timescale) with $M_\mathrm{ej} = 0.62~ \mathrm{M}_\odot$ and $v_\mathrm{ej} = 10880$~km~s$^{-1}$. 

A complementary constraint on the transparency time comes from persistent radio sources (PRSs). The association of a subset of FRBs to compact PRS provides evidence that at least some FRBs originate within dense plasma environments, consistent with a young, flaring magnetar embedded in a SN remnant as the central engine \citep{Chatterjee17, Niu22, Bruni24, Bruni2024b}. In this picture, the dense SN ejecta first expands outward and becomes transparent to MHz radio emission. The interaction between the SN blastwave and the surrounding circum-burst medium drives shocks that produce synchrotron radio emission, observed as persistent radio sources \citep{Margalit18, Zhao21}. Indeed, modeling based on the observed compact sizes and luminosities of PRSs associated with these FRBs offers an independent constraint on the progenitor age, with estimated ages of $\sim$10–40 years \citep{Margalit19, Zhao21, Bhattacharya24}. This range inferred from the PRS associations is characteristic of the expected transparency time, and we therefore adopt an upper limit of 10 years for the minimum transparency time.
 
Moreover, we emphasize that the transparency time range is not absolute and only provided here for illustrative purposes. Accordingly, FRBs may still be detectable from even younger SNe. Asymmetries in CCSNe are commonly observed, as evidenced by the morphology of nearby SN remnants and the intrinsic polarization properties of CCSNe \citep{Larsson13, Reilly16, Tinyanont21, Milisavljevic24}. This implies that in extreme cases, the transparency timescale could in principle be zero if the FRB is beamed along a direction largely free of SN ejecta. Conversely, in systems with higher ejecta masses, lower ejecta velocities, or less favorable geometry, the minimum transparency timescale could significantly exceed the range shown in Figure \ref{fig:dd_cdf}.

Now, we compare the times since SN discovery to the plausible minimum transparency timescales for our simulated FRBs. If we assume that it takes at least 10~yr for FRBs to become observable from these systems, we find that out of all our positionally-matched transient, 5$\%$ would have had an environment transparent enough for the FRB to escape. On the other hand, 23$\%$ of all {\it classified} SNe and 32$\%$ of all CCSNe were discovered more than 10 years ago. These fractions correspond to $\approx$ 4700 classified SNe with an average redshift of $z=0.017$ and $\approx$ 1700 CCSNe with an average redshift of $z=0.015$. Among the classified SNe, the fraction of SNe Ia and CCSNe make up approximately 59.2$\%$ and 35.8$\%$, respectively. The fact that only one quarter of the classified SN population have a long discovery timescale ($>$10 yr) means that an FRB matched with such an SN would be even more significant than described in Section \ref{sec:Pcc}.

If we assume a transparency timescale of 6.4~yr (the derived lower limit on the minimum transparency timescale), we find that up to $\approx 7\%$ of the matched optical transients could be associated with a detectable FRB, given the opacity of the SN ejecta and assuming that these transients are all SNe. In comparison, $\approx 30\%$ of all cataloged SNe and $\approx 40\%$ of CCSNe were discovered more than 6.4 years ago, corresponding to the expected fraction of SNe that could be associated with a detectable FRB if every SN were to have such an association. Among the classified SNe, 58.8$\%$ are SNe Ia and 41.2$\%$ are CCSNe. These fractions reinforce the value of continued searches for past SNe, which may be especially fruitful for any population of older, nearby transients that emit detectable FRBs.

Beyond past SNe, we have purposely developed this machinery such that it can be applied to optical emission associated with FRBs across a broad range of luminosities, durations, and delay times relative to the FRB \citep{Yang19, Chen20, Zhang2023}. For instance, in the scenario in which an FRB excites an afterglow in the plasma surrounding a magnetar, it is possible to produce a prompt bright ($\gtrsim$10$^{41}$ erg s$^{-1}$) optical flash on millisecond to $\sim$\,1~s timescales, or a faint ($\lesssim$10$^{39}$ erg s$^{-1}$) optical afterglow lasting seconds to minutes \citep{Metzger19, Beloborodov20}. On month to year timescales, the hypernebula model that involves a stellar-mass compact object accreting at super-Eddington rates from a stellar companion predicts a faint optical counterpart from the shock-ionized plasma or reprocessing of beamed X-rays from ultra-luminous X-ray jets \citep{Sridhar22}. As such, this machinery makes it possible to identify associations with such phenomena, as long as they are cataloged in TNS.

\subsection{Redshift Distribution Comparisons} \label{redshifts}

\begin{figure}
    \centering
    \includegraphics[width=\linewidth]{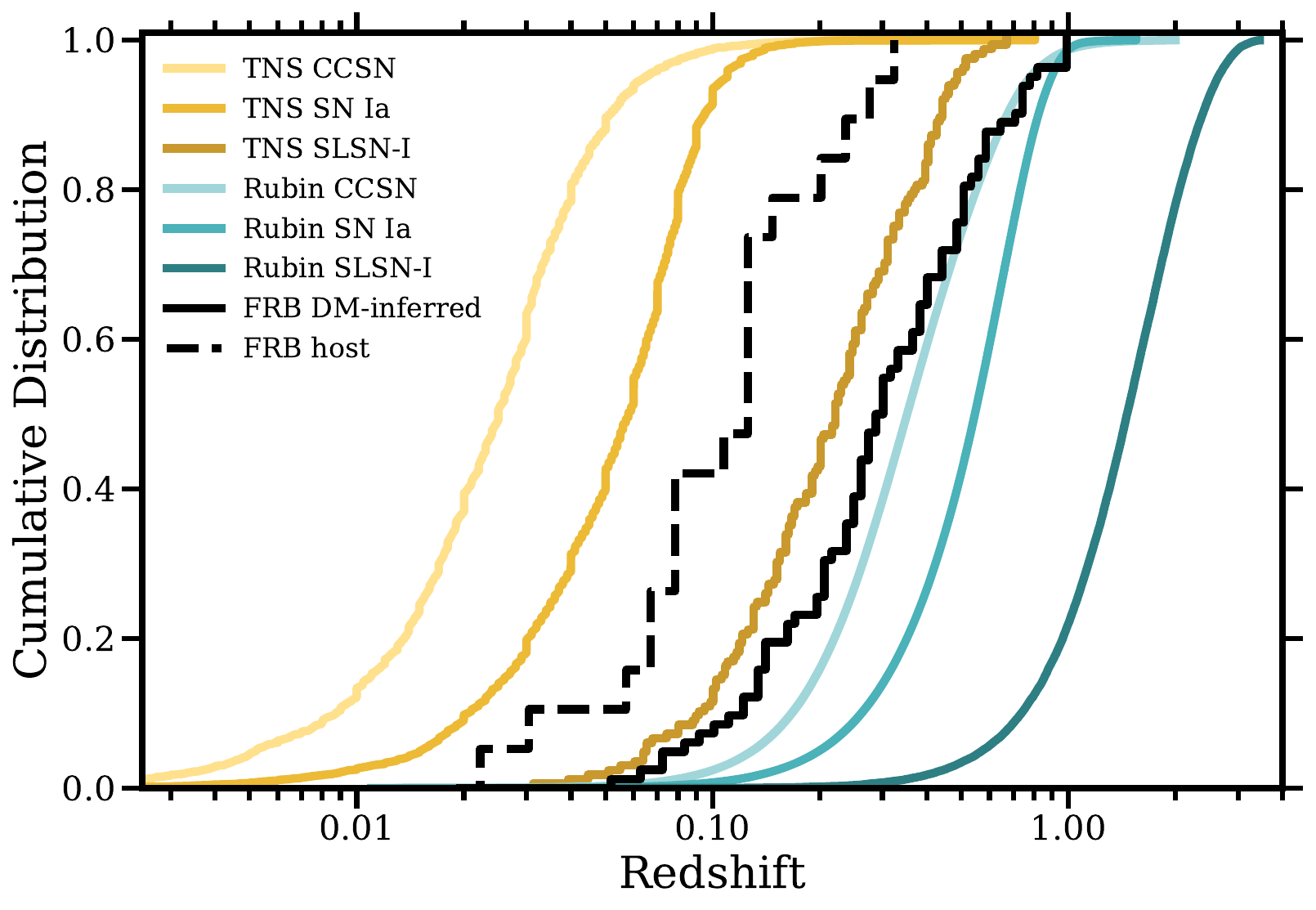}
    \caption{CDFs of redshift for the CHIME VLBI-localized FRBs alongside CCSN, SN Ia, and SLSN-I. The distributions for CCSNe, SNe Ia, and SLSNe-I from TNS are shown as different shades of yellow. Median redshifts inferred from CHIME-KKO FRB DMs are indicated by a solid black line, while a subset of them with spectroscopic redshifts are shown as dashed lines. Projecting forward, redshift distributions of simulated SNe that would be observed by the Rubin Observatory from the \texttt{PLAsTiCC} dataset \citep{Kessler19} are shown in blue. Although the current population of SNe in TNS, except SLSNe-I, is concentrated at $z<$0.1, next-generation optical surveys such as the Rubin Observatory will be able to detect SNe out to higher redshifts ($z<$1), overlapping with the redshift range where most FRBs are observed, and even farther ($z\sim$3.5) in the case of SLSNe-I.}
    \label{fig:forecast}
\end{figure}

To compare the volumes (and thus detectability) probed by CHIME-KKO FRBs\footnote{It is expected that the CHIME-KKO FRB sample will have a similar redshift distribution to that detected by the full Outrigger array, and thus any statements that apply to CHIME-KKO FRBs here can be broadly applied.}, SNe reported in TNS, and optical transients soon to be detected by the Rubin Observatory, we plot their known and expected redshift distributions in Figure~\ref{fig:forecast}. For FRBs, their observed redshift distribution is inherently biased towards bright galaxies and FRBs with localizations precise enough for robust host associations (explained further later in this section). For SNe, their redshift determinations and spectral classifications require them to be sufficiently luminous or nearby, limiting the volume for the known population of spectroscopically classified SNe in TNS.  On the other hand, extragalactic FRBs have been observed at redshifts up to $z \sim$ 1 \citep{Ryder23,Connor24}, and thus the volume in which FRB-SN associations can be made is primarily limited by the detectability and spectral classification of SNe. The majority of classified SNe are at $z<0.05$ for CCSNe \citep{Perley20}, whereas the rarer and more luminous SLSNe are generally detected at higher redshifts at $z<0.3$ \citep{Gomez24}. We include SLSNe in this section as their shared host properties to the first repeating FRB\,20121102A and their similar inferred magnetar engine properties \citep{Metzger17,Margalit18} have motivated searches for PRS and FRB emission from SLSNe years after the SN discovery \citep{law19,Eftekhari21}.

As described in Section~\ref{sec:redshift_frb}, we infer the probability distribution of redshift for each FRB from its DM, and determine the median value for each CHIME-KKO FRB (from \citealt{KKOhost25}). For FRBs without known redshift, we plot the resulting CDF from the median values in Figure~\ref{fig:forecast}. We also include spectroscopic redshifts from \cite{KKOhost25} for 19 CHIME-KKO FRBs with robust host associations. We find a median DM-inferred redshift of $z = 0.28$, which is slightly lower than the results from \citet{Shin23}, where the predicted redshift distribution for the CHIME/FRB sample peaks at $z \approx 0.36$. While it is reasonable to assume that the CHIME/FRB Outriggers redshift distribution has the same peak as that of CHIME/FRB, future work is needed to characterize the potential biases that may affect the redshift distribution of the Outrigger samples compared to CHIME/FRB.

On the other hand, the median spectroscopic redshift ($z_\mathrm{spec}$ = 0.1) of the CHIME-KKO FRBs is notably lower than the distribution of those with only DM-inferred redshifts. This can be naturally explained by a combination of two reasons. First, while DM can be used as a distance proxy, especially where it is dominated by DM$_{\rm cosmic}$ at high redshift, the assumptions made in the DM-redshift calculation include significant systematic uncertainties that produce more scatter than direct redshift measurements. A striking example of DM excess is FRB\,20190520B whose true redshift is much lower than what is inferred from its DM because of additional contributions from the host galaxy and foreground galaxy clusters along the line of sight \citep{Ocker22, Lee23}. This highlights the limitations of using DM alone as a distance estimator, particularly in cases of significant excess as the redshifts of these sources will be biased high in the DM-inferred redshift cases. However, we note that the joint probability distribution of the redshift and DM$_\mathrm{EG}$ used in this work still encompasses the true redshift of FRB\,20190520B, despite its large DM excess.

Second, the requirement of a robust host association based on archival optical survey data imposes a limiting magnitude on any identified host, introducing a selection bias in which FRBs with faint (i.e., low luminosity and/or high-redshift) hosts will not have identifiable host associations and thus no redshift estimates. CHIME-KKO FRBs with confidently associated hosts reported thus far are likely biased toward bright, low-redshift galaxies; this is also reflected in their systematically lower DMs compared to the full population \citep{KKOhost25}. This follow-up selection bias contributes to the observed difference in the spectroscopic and DM-inferred redshift distributions of CHIME-KKO FRBs \footnote{For CHIME-KKO FRBs with $z_\mathrm{spec}$, we compare to their DM-inferred redshift estimates and find that the latter often overestimate the true redshift, which means they should be treated more as upper limits.}.

% PARAGRAPH ABOUT RUBIN and TNS transient z 
To illustrate the relative detectability of optical transients, we plot CDFs of redshift for SNe\,Ia, CCSNe, and SLSNe-I from TNS in Figure~\ref{fig:forecast} to compare the volumes that they probe with those of the FRBs. We find that nearly all TNS SNe are detected within a volume of $z<0.3$; specifically, $>$99$\%$ of Type\,Ia SNe and CCSNe and 70$\%$ of SLSNe are at or below this redshift. Among CHIME-KKO FRBs with spectroscopic redshifts, 95$\%$ fall within this volume. This implies that if every CHIME-KKO FRB with a spectroscopic redshift has a detectable and classified SN counterpart, we should expect to observe an associated SN in existing optical surveys for almost all of them. When considering CHIME-KKO FRBs with DM-inferred redshifts instead, the expected fraction of TNS SNe decreases to 55$\%$.

For CCSNe and SNe Ia, Figure~\ref{fig:forecast} shows that FRBs at $z>0.1$ are mostly beyond the reach of existing SN surveys. As a result, FRB-SN associations at $z<0.05$ will provide the most promising path to test the hypothesis that FRBs originate from magnetars formed via the core collapse of massive stars or a more exotic channel such as the accretion-induced collapse of a WD \citep{Fryer99} or a neutron star merger \citep{Giacomazzo13}. In order to assess the significance of such an association in the near future, we estimated the waiting time for a chance coincidence from the $P_{cc}$ curve in Section~\ref{sec:Pcc}. If we find more than one unambiguous match in the next few years, they would strongly signify a real association between SNe and at least some FRBs and provide a definitive test of these progenitor channels. To this end, we stress the need to search for past SNe associated with FRBs at $z\lesssim0.05$ where cataloged, spectroscopically-classified SNe are relatively complete.

Furthermore, the dynamic, magnetized environments around PRS-associated FRBs can drive variations in DM and Faraday rotation measure (RM) \citep{Yang17, Piro18}. A match involving large DM and RM variations in FRB properties and an SN could be especially compelling. In contrast, the current sample of SLSNe extends out to $z \approx 0.7$, a range that encompasses all CHIME-KKO FRBs with $z_\mathrm{spec}$. This demonstrates that although SLSNe are intrinsically rarer than CCSNe, potential associations with SLSNe can be probed to significantly higher redshifts (larger volumetric overlap).

An illustrative example is the recent discovery of FRB\,20250316A in the extremely nearby ($z=0.006$) galaxy NGC\,4141 \citep{Ng2025ATel}. This galaxy also hosted two known past SNe, the Type II SNe\,2008X \citep{Boles08} and 2009E \citep{Bole09}, although the CHIME-KKO localization confirms that the FRB is not positionally coincident with either of these SNe \citep{Leuong25ATel} or any other known, cataloged optical transient. However, its proximity, extremely precise localization, and occurrence in a prolific SN-producing galaxy demonstrates a promising opportunity to search for an associated, evolved SN that may have been missed by previous time-domain surveys for other FRBs in the future.

Finally, the Rubin Observatory will vastly expand the volume and number of optical SN detections, and so we compare the FRB redshift distributions with those expected for Rubin SNe. We acknowledge that the overlap between the Rubin and CHIME footprints is not favorable for FRB-SN matches, but the following discussion is simply illustrative. For SNe that would be observed by Rubin through LSST, we used a simulated dataset for the Photometric LSST Astronomical Time Series Classification Challenge (PLAsTiCC), which consists of optical transients expected to be discovered by Rubin under realistic observing conditions over a 3-yr period \citep{Kessler19}. With the depth and wide-field coverage of Rubin/LSST \citep{LSST} on the horizon, the median redshifts of CCSNe, SNe Ia, and SLSNe are extended to $z = 0.35$, $z = 0.55$ and $z = 1.46$, respectively. In comparison, the current TNS samples reaches only $z = 0.03$, $z = 0.06$ and $z = 0.22$ for the same SN types. Moreover, Figure \ref{fig:forecast} shows that the volume probed by those with spectroscopic redshifts is fully contained within that of Rubin SNe, and we will not be as limited by the detectability volume of SNe.

Within the redshift range of $z < 1$, where virtually all FRBs are discovered and encapsulating all of the DM-inferred redshifts of our CHIME-KKO FRB sample (which again, can serve as a proxy for the DM-inferred redshift distribution of bursts detected by the full Outrigger array), this volume also contains $99\%$ of simulated Rubin CCSNe and SNe Ia (although only 22\% of SLSNe, which are detected much farther). This increase in overlap between FRBs and the two main classes of SNe compared with the existing TNS SN population suggests that if FRB-SN associations exist, the rate at which they are found in (and beyond) the Rubin era will only increase, even as FRB detectability pushes to higher redshifts with DSA-2000, CHORD, and SKA on the horizon \citep{SKA, SKA_mac, CHORD}. Indeed, under the assumptions that each new FRB comes from a known, past SN, we can expect considerably more associations, particularly at higher redshifts, in the future. Moreover, as SNe are discovered and become transparent to FRBs throughout the 10~yr survey of LSST, the incidence of FRB-SN detection would also increase. Assuming cosmological parameters from \cite{planck2018}, this increase could reach a factor of $\sim$~300 even ignoring the steady growth in FRB detection rates from new experiments that will probe the LSST volume. This is a lower limit, as the volumetric SN rate increases with redshift, implying a larger population of magnetar progenitors capable of producing FRBs. One caveat is that the years-to-decade delay between SNe and FRBs means the discovery rate may initially be modest but is expected to grow steadily in the latter part of LSST's 10~yr survey and beyond.

\section{Conclusions} \label{sec:conclusions}
We have performed systematic searches for positional and redshift coincidences of CHIME-KKO, as well as all well-localized literature FRBs, with optical transients in TNS. The novel machinery developed in this work provides a direct test of the magnetar progenitor model of FRBs for sources originating in core-collapse or more exotic magnetar formation channels that are expected to produce optical transients \citep[e.g.,][]{Fryer99,Giacomazzo13}. Our main results are summarized as follows:

\begin{itemize}
    % Null results: no matches across all FRBs
    \item We do not identify any statistically significant positional or redshift associations between 83 CHIME-KKO or 93 well-localized literature FRBs with all optical transients in the TNS. The sole exception is the previously suggested association between FRB\,20180916B and AT\,2020hur \citep{Li22}, demonstrating the performance of our crossmatching pipeline. However, given the limited number of detections and ancillary data for this transient, we cannot comment further on the robustness of this association.
    \item While our work primarily focuses on optical transients that occurred before the FRB discovery (i.e., historical transients), we also find no positionally coincident cataloged transient that occurred {\it after} any known well-localized FRB (beyond FRB\,20180916B).
    % Pcc + % chance match rate prediction
    \item We simulate a population of FRBs based on parameters of the CHIME-KKO sample to determine the probability of chance coincidence ($P_{cc}$) of FRB-transient matches. In the full CHIME/FRB Outrigger era, we expect one chance coincidence for every $\sim$~22,700 subarcsecond-localized FRBs ($\sim $30–60~yr given CHIME/FRB Outrigger rates). Thus, any near-future match between an FRB and classified SN will likely be a physical association. 
    % Ejecta transparency
    \item We derive a minimum transparency timescale between the SN explosion and the first observable FRB through the SN ejecta, finding a range of reasonable minimum timescales of $\gtrsim$6.4--10~yr. We estimate that 5--7$\%$ of the transients matched to simulated FRBs are older than the transparency timescale, while 23--30$\%$ of all cataloged SNe and 32--41$\%$ of CCSNe are currently old enough to have detectable FRB emission.
    % Redshift comparisons current
    \item 95$\%$ of CHIME-KKO FRBs with spectroscopic redshifts fall within a volume of $z < 0.3$, where most classified SNe are detected by current surveys. If every such FRB has an associated SN, we expect to eventually observe a detectable counterpart for almost all of them. 
    % Redshift Forecast
    \item In the next year, Rubin will dramatically increase both the number of known SNe and the volume over which they can be detected by multiple orders of magnitude.  The Rubin survey volume for optical transients will better match the volume probed by all planned and near-future FRB experiment upgrades, including the CHIME/FRB Outriggers, compared with known SNe from current surveys. Thus, the expected number of FRB-SN associations will increase toward redshifts $z \sim 1$ (below which nearly all FRBs are discovered so far). This improved redshift overlap will significantly increase the rate of expected FRB-SN associations (if they exist), even as FRB detectability extends to higher redshifts.
    % Versatiility of this pipeline
    \item The machinery introduced here was specifically applied to CHIME-KKO FRBs, literature FRBs, and TNS transients, but it is equally suitable for any FRB-optical transient associations and can be easily adapted for different FRB experiments. More broadly, it is inherently versatile and can be applied to any type of transient, in time and over wavelength. To aid such searches, we have made the code publicly available.
\end{itemize}

As we enter the era of hundreds to thousands of VLBI-localized FRBs, systematic searches for past SNe as optical counterparts offer a direct and powerful test of the magnetar progenitor model. We encourage the community to build on the tools provided here and apply them to FRBs from any experiment to search for cross-matches between various classes of transients in real time, in particular targeting nearby, older SNe, where associated FRB emission may already be detectable. If FRBs are eventually identified in future searches from past SNe, such associations could offer a new avenue to better constrain the FRB beaming fraction. Beyond optical transients, we also encourage positional coincidence searches to slow-evolving emission such SN remnants, particularly for the nearest FRBs as most known extragalactic remnants are located within a few tens of megaparsecs. Finally, the Rubin Observatory will also unleash an unprecedented number of new SNe, and should particularly motivate FRB experiments with larger footprint overlap to carry out such searches.

\vspace{-3mm}
\begin{acknowledgments}

The authors are grateful for valuable discussions regarding supernovae with Steve Schulze and TNS querying with Ofer Yaron. Y.D. is supported by the National Science Foundation (NSF) Graduate Research Fellowship under grant No. DGE-2234667. C.D.K. gratefully acknowledges support from the NSF through AST-2432037, the HST Guest Observer Program through HST-SNAP-17070 and HST-GO-17706, and from JWST Archival Research through JWST-AR-6241 and JWST-AR-5441. W.F. gratefully acknowledges support by the NSF under grant no. AST-2206494 and CAREER grant No. AST-2047919, the David and Lucile Packard Foundation, the Alfred P. Sloan Foundation, and the Research Corporation for Science Advancement through Cottrell Scholar Award \#28284. B.\,C.\,A. is supported by an Fonds de Recherche du Quebec—Nature et Technologies~(FRQNT) Doctoral Research Award. A.P. is funded by the NSERC Canada Graduate Scholarships -- Doctoral program. V.S. is supported by a FRQNT Doctoral Research Award. A.M.C. is a Banting Postdoctoral Researcher. A.B.P. is a Banting Fellow, a McGill Space Institute~(MSI) Fellow, and a FRQNT postdoctoral fellow. D.M. acknowledges support from the French government under the France 2030 investment plan, as part of the Initiative d'Excellence d'Aix-Marseille Universit\'e -- A*MIDEX (AMX-23-CEI-088). M.W.S. acknowledges support from the Trottier Space Institute Fellowship program. E.F. is supported by the NSF under grant number AST-2407399. A.P.C. is a Vanier Canada Graduate Scholar. K.S. is supported by the NSF Graduate Research Fellowship Program. V.M.K. holds the Lorne Trottier Chair in Astrophysics $\&$ Cosmology, a Distinguished James McGill Professorship, and receives support from an NSERC Discovery grant (RGPIN 228738-13). K.W.M. holds the Adam J. Burgasser Chair in Astrophysics and is supported by NSF grant 2018490. K.N. is an MIT Kavli Fellow. P.S. acknowledges the support of an NSERC Discovery Grant (RGPIN-2024-06266). C. L. acknowledges support from the Miller Institute for Basic Research at UC Berkeley.

The Fast and Fortunate for FRB Follow-up team acknowledges support from NSF grants AST-1911140, AST-1910471, and AST-2206490. 
We acknowledge that CHIME and the \kkoname{} Outrigger (KKO) are built on the traditional, ancestral, and unceded territory of the Syilx Okanagan people. \kkoname{} is situated on land leased from the Imperial Metals Corporation. CHIME operations are funded by a grant from the NSERC Alliance Program and by support from McGill University, University of British Columbia, and University of Toronto. CHIME/FRB Outriggers are funded by a grant from the Gordon \& Betty Moore Foundation. We are grateful to Robert Kirshner for early support and encouragement of the CHIME/FRB Outriggers Project, and to Dusan Pejakovic of the Moore Foundation for continued support. The CHIME/FRB Project was funded by a grant from the CFI 2015 Innovation Fund (Project 33213) and by contributions from the provinces of British Columbia and Québec, and by the Dunlap Institute for Astronomy and Astrophysics at the University of Toronto. Additional support was provided by the Canadian Institute for Advanced Research (CIFAR), the Trottier Space Institute at McGill University, and the University of British Columbia. This research has made use of the NASA/IPAC Extragalactic Database (NED), which is funded by the National Aeronautics and Space Administration and operated by the California Institute of Technology.

\end{acknowledgments}

\vspace{-0.2mm}

\software{
{\tt astropy} \citep{astropy13, astropy18, astropy22},
{\tt FFFF-PZ} \citep{Coulter2022, Coulter2023},
{\tt matplotlib} \citep{Matplotlib},
{\tt numpy} \citep{numpy}, 
{\tt pandas} \citep{reback2020pandas}, 
{\tt SAOImageDS9} \citep{DS9},
{\tt scipy} \citep{Scipy}, 
{\tt shapely} \citep{Shapely}
}

%\clearpage
\bibliographystyle{aasjournal}
\bibliography{references}

\end{CJK*}
\end{document}